\documentclass[sigconf]{acmart} %% CCS: DO NOT REMOVE
\AtBeginDocument{%
  }

\setcopyright{none}
\acmConference[Preprint]{Preprint}{2026}{}
\renewcommand\footnotetextcopyrightpermission[1]{}

\usepackage{graphicx}
\usepackage{xspace}
\usepackage{textcomp}
\usepackage{url}
\usepackage{array}
\usepackage{makecell}
\usepackage{amsmath}  % Necesario para \mathcal
\usepackage{algorithm}
\usepackage{algpseudocode}
\usepackage{multirow, colortbl, xcolor, booktabs}
\usepackage{svg}
\usepackage{booktabs} % Para tablas con buen estilo
\usepackage{xcolor}
\usepackage[most]{tcolorbox}
\usepackage{tcolorbox}
\usepackage{listings}
\usepackage{caption}
\usepackage{subcaption}
\usepackage{xcolor}
\usepackage{adjustbox}
\definecolor{darkgreen}{RGB}{0,120,0}
\newcolumntype{C}[1]{>{\centering\arraybackslash}p{#1}}

\usepackage{tikz}
\newcommand{\circledblack}[1]{%
  \tikz[baseline=(char.base)]{
    \node[shape=circle, fill=black, text=white, inner sep=0.6pt, font=\bfseries] (char) {#1};
  }%
}
\newcommand{\circled}[1]{%
  \tikz[baseline=(char.base)]{
    \node[shape=circle,draw,inner sep=1pt] (char) {\small #1};}}

\usepackage{tikz}
\usepackage{amsmath}

\usepackage{filecontents}
\newcommand{\tool}{\textsc{FunFuzz}\xspace}
\begin{document}

%%
%% The "title" command has an optional parameter,
%% allowing the author to define a "short title" to be used in page
%% headers.

\title{\tool: An LLM-Powered Evolutionary Fuzzing Framework} %% CCS: you MUST provide a title

%%
%% The "author" command and its associated commands are used to define
%% the authors and their affiliations.
%% Of note is the shared affiliation of the first two authors, and the
%% "authornote" and "authornotemark" commands
%% used to denote shared contribution to the research.

%% CCS: at submission time, the submission MUST be anonymized. Hence
%% authors MUST be commented out.

\author{Mario Rodríguez Béjar}
\email{mario.rodriguezb1@um.es}
\affiliation{%
    \institution{Universidad de Murcia}
    \city{Murcia}
    \country{Spain}
}

\author{Bernardino Romera Paredes}
\email{bernardino@hiverge.ai}
\affiliation{%
    \institution{Hiverge}
    \city{London}
    \country{England}
}

\author{Jose Luis Hernández-Ramos}
\email{jluis.hernandez@um.es}
\affiliation{%
    \institution{Universidad de Murcia}
    \city{Murcia}
    \country{Spain}
}

%%
%% By default, the full list of authors will be used in the page
%% headers. Often, this list is too long, and will overlap
%% other information printed in the page headers. This command allows
%% the author to define a more concise list
%% of authors' names for this purpose.
%\renewcommand{\shortauthors}{Trovato et al.}
\renewcommand{\shortauthors}{Rodríguez Béjar et al.}
%%
%% The abstract is a short summary of the work to be presented in the
%% article.
\begin{abstract} %% CCS: an abstract MUST be provided.
  
  Modern fuzzers increasingly use Large Language Models (LLMs) to generate structured inputs, but LLM-driven fuzzing is sensitive to prompt initialization and sampling variance, which can reduce exploration efficiency and lead to redundant inputs. We present \tool, a multi-island evolutionary fuzzing framework that runs several isolated searches in parallel and periodically migrates high-value candidates to maintain diversity. \tool derives initial generation prompts from documentation and initializes islands with topic-specific instructions, then continuously adapts prompts using feedback-guided selection. During fuzzing, candidates are prioritized by incremental compiler coverage, while compiler-internal failure signals are used to identify crash-inducing inputs. We evaluate \tool on compiler fuzzing, where inputs are source programs and success is measured by compiler coverage and unique compiler-internal failures. Across repeated 24-hour campaigns on GCC and Clang, \tool achieves higher compiler coverage than previous LLM-driven baselines and discovers more unique failure-triggering inputs.
\end{abstract}

%%
%% The code below is generated by the tool at http://dl.acm.org/ccs.cfm.
%% Please copy and paste the code instead of the example below.
%%%
%\begin{CCSXML}
%<ccs2012>
%<concept>
%<concept_id>10011007.10011074.10011099.10011102.10011103</concept_id>
%<concept_desc>Software and its engineering~Software testing and debugging</concept_desc>
%<concept_significance>500</concept_significance>
%</concept>
%<concept>
%<concept_id>10010147.10010178.10010205</concept_id>
%<concept_desc>Computing methodologies~Search methodologies</concept_desc>
%<concept_significance>300</concept_significance>
%</concept>
%<concept>
%<concept_id>10011007.10010940.10010992.10010998.10011001</concept_id>
%<concept_desc>Software and its engineering~Dynamic analysis</concept_desc>
%<concept_significance>100</concept_significance>
%</concept>
%</ccs2012>
%\end{CCSXML}

%\ccsdesc[500]{Software and its engineering~Software testing and debugging}
%\ccsdesc[300]{Computing methodologies~Search methodologies}
%\ccsdesc[100]{Software and its engineering~Dynamic analysis}
%%
%% Keywords. The author(s) should pick words that accurately describe
%% the work being presented. Separate the keywords with commas.
 %% CCS: DO NOT REMOVE but you MAY update

% \received{20 February 2007} 
% \received[revised]{12 March 2009}
% \received[accepted]{5 June 2009}

%%
%% This command processes the author and affiliation and title
%% information and builds the first part of the formatted document.
\maketitle

\section{INTRODUCTION}\label{sec:introduction}

Language-processing systems such as compilers, interpreters, and runtime engines underpin modern software development and toolchains. Defects in these components can propagate widely, affecting downstream applications and the broader software supply chain. Over the past decades, \textit{fuzzing} has proven highly effective at uncovering real bugs in complex systems at scale~\cite{mallissery2023demystify}. In particular, compiler fuzzing remains challenging because the input space is highly structured and semantically rich: deep compiler behavior often requires programs that combine multiple language features and non-trivial interactions. As a result, campaigns can converge to shallow patterns that are easy to generate while missing rare combinations that exercise deeper compilation pipelines \cite{ma2023survey}. Modern compiler evolution further amplifies the problem because new features and optimizations reshape the reachable behavior space and invalidate assumptions embedded in existing strategies \cite{kwon2025optimization}.

One recurring challenge lies in the high degree of specialization required to adapt these tools to a specific system under test. Developing a fuzzer for a new compiler or language variant often involves significant manual engineering, from designing intricate grammar rules to implementing target-specific heuristics~\cite{yang2011finding}. This specialization reduces portability, as techniques fine-tuned for one domain may fail to deliver similar results when applied in a different context. As a result, re-targeting to new language variants or new compiler subsystems often requires substantial effort and expertise~\cite{holler2012fuzzing}. Another limitation concerns adaptability: modern compilers evolve quickly, adding new features and optimizations that render previously effective fuzzing strategies less potent. Previous studies~\cite{even2022csmithedge} show that even long-standing tools like Csmith, once highly effective, struggle to uncover new defects in the latest versions of GCC and Clang.
A third limitation is sustained semantic diversity. Many approaches bias exploration toward a tractable subset of language constructs or toward neighborhoods around a fixed seed corpus, which leaves rare feature interactions under-exercised~\cite{herrera2021seed}. Together, these limitations motivate fuzzing techniques that reduce target-specific engineering, remain effective as compilers evolve, and maintain exploration pressure toward semantically rich programs rather than converging to shallow patterns.

Large language models (LLMs) \cite{zhu2025software} offer a promising alternative to manual generators and corpus-heavy fuzzing because they can synthesize source programs that are syntactically well-formed and semantically varied with limited target-specific engineering~\cite{huang2025challenges}. Prior LLM-assisted fuzzers demonstrate that prompt-guided generation can bootstrap fuzzing on complex targets where traditional input generators are expensive to build~\cite{xia2024fuzz4all}. However, LLM-driven fuzzing faces a practical gap: generation can collapse to repetitive program styles, early samples can bias the trajectory of a campaign, and exploration often stagnates without an explicit mechanism that preserves diversity while still exploiting useful feedback. This motivates a search procedure that remains robust under LLM stochasticity and sustains long-term exploration pressure toward new compiler behaviors rather than repeatedly refining a narrow set of patterns.

Existing LLM-assisted fuzzers such as Fuzz4All rely on prompt distillation to initialize generation, but typically start from a single shared prompt, which can bias the search trajectory early and limit semantic diversity across generated programs. In contrast, \tool introduces per-island seed instructions, enabling multiple semantically distinct starting points that encourage early divergence and reduce prompt-induced bias.

Despite recent progress in LLM-based compiler fuzzing, existing pipelines still rely largely on implicit or coarse-grained feedback to guide search. In Fuzz4All, generation-time decisions are primarily tied to validity and execution outcomes~\cite{xia2024fuzz4all}, while Kitten follows a syntax-guided mutation workflow (parse, mutate, compile, check) and uses abnormal compiler behaviors (e.g., crashes and hangs) as bug oracles~\cite{xie2025kitten}. Although these signals are effective for filtering candidates, they provide limited explicit guidance about how much new compiler behavior a test actually explores, so search may drift toward repeatedly producing superficially valid yet structurally similar inputs. To address this limitation, \tool introduces an explicit fitness objective for candidate ranking. Importantly, this objective is not based on raw coverage alone; it is a richer fitness signal built from results obtained under multiple compilation configurations, so selection reflects more robust behavioral novelty across settings. In parallel, because compiler fuzzing benefits from sustained diversity rather than convergence to a single “best” artifact, \tool combines this fitness-guided selection with a multi-island evolutionary process that maintains partially independent search trajectories. Together, these components provide both direction (via explicit fitness) and diversity (via island-level separation), forming the basis of the framework presented next.

\textbf{Our work.} We present \tool, an evolutionary fuzzing framework that integrates LLM-based program synthesis with a multi-island search procedure. \tool starts from an autoprompting-style distillation step that turns user-provided documentation and examples into a base prompt~\cite{xia2024fuzz4all}. Building on these ideas, \tool derives a set of per-island seed instructions at initialization. Each island combines the shared base prompt with a different seed instruction, which creates multiple semantically distinct starting points. During fuzzing, each island maintains an independent population and prioritizes candidates using coverage-guided fitness computed on an instrumented compiler. Islands evolve independently in parallel during local search, and a periodic migration transfers a small fraction of high-value candidates from stronger islands to weaker ones without resetting entire populations. This design aims to sustain exploration under LLM stochasticity: per-island initialization encourages early divergence, local feedback preserves independent search frontiers, and migration mitigates stagnation while avoiding premature homogenization.

We evaluate \tool on \emph{compiler fuzzing} for C and C++, where the inputs are LLM-generated source programs and the goal is to trigger crashes and internal compiler failures in modern compilers such as GCC and Clang. Our evaluation targets modern versions of GCC and Clang under long-running 24-hour campaigns and compares against representative baselines: the LLM-assisted evolutionary fuzzer Fuzz4All~\cite{xia2024fuzz4all} and the high-throughput mutational fuzzer Kitten (C only)~\cite{xie2025kitten}. On GCC-based targets, \tool reaches up to +27.1\% coverage over standard Fuzz4All in C and up to +31.0\% in C++. On Clang-based targets, \tool reaches up to +9.3\% in C and up to +4.4\% in C++. Overall, \tool discovers 119 unique compiler bugs, of which 80 were confirmed by developers. We further report controlled ablations that isolate the contribution of multi-island search, prompt initialization, and scoring components. While \tool is designed to be agnostic to the system under test, our evaluation is limited to C/C++ compilers; validating its effectiveness on other structured-input domains (e.g., interpreters or protocol parsers) remains future work.

\paragraph{Contributions.} We make the following contributions:

\begin{itemize}
    \item \textbf{Multi-island LLM-assisted evolutionary fuzzing.} We introduce a multi-island fuzzing loop that maintains independent populations with local feedback and uses periodic migration to mitigate stagnation without discarding search context.

    \item \textbf{Prompt distillation with per-island semantic initialization.} We extend autoprompting-style distillation with per-island seed instructions that create multiple semantically distinct starting points while preserving syntactic validity.

    \item \textbf{Evaluation on modern GCC and Clang.} We provide a 24-hour evaluation for C/C++ that measures compiler coverage and bug-finding performance against strong baselines, plus ablations that quantify the impact of core design choices.
\end{itemize}

\section{Related Work}

Previous work related to \tool spans (i) compiler fuzzers that emphasize high-throughput mutation and curated corpora, (ii) LLM-based approaches that synthesize structured inputs from documentation, and (iii) evolutionary search methods that preserve diversity through multi-population exploration. We summarize these lines of work and then position \tool relative to the closest baselines.

\textbf{Structured-input and compiler fuzzing}. Coverage-guided fuzzing is a widely used approach for scalable bug discovery in complex systems, with well-known engines such as American Fuzzy Lop (AFL) \cite{afl}, AFL++ \cite{fioraldi2020afl++}, and libFuzzer \cite{libfuzzer}. Compiler testing adds an additional challenge: inputs are programs with rich syntactic and semantic constraints, and techniques must trade validity, diversity, and throughput \cite{ma2023survey}. Generation-based tools such as Csmith \cite{yang2011finding} provide well-defined randomized programs and have historically exposed many compiler defects, but can saturate over time and require maintenance as languages and compiler behaviors evolve \cite{ma2023survey}. Mutation-heavy compiler fuzzers focus on throughput and seed exploitation; recent systems such as Kitten \cite{xie2025kitten} represent a strong baseline in this space, using curated corpora with aggressive mutations to maximize coverage over modern toolchains. These approaches scale well, but they often depend on the semantic reach of seeds and mutation operators, which can leave rare feature interactions under-exercised in large and evolving language front-ends.

\textbf{LLM-assisted fuzzing for structured inputs}. Recent work shows that LLMs can synthesize structured inputs directly from documentation and examples, reducing the need for hand-written grammars and domain-specific generators \cite{deng2023large, huang2024large}. This direction is particularly attractive for language-processing targets, where surface validity alone is insufficient and deeper behavior often requires non-trivial compositions of language features \cite{ma2023survey}. Fuzz4All \cite{xia2024fuzz4all} is the closest previous system to ours in spirit: it distills documentation into prompts and uses LLM generation inside a feedback-guided loop, which enables cross-domain fuzzing without target-specific grammars. Other systems explore complementary roles for LLMs, including LLM-guided transformations over existing seeds or hybrid pipelines that mix semantic edits with lower-level mutations \cite{zhang2024llamafuzz}. For compiler fuzzing specifically, recent work also uses LLMs to improve mutation quality and diversify transformations, which supports the premise that model-driven synthesis can uncover defects that corpus-driven mutation may miss \cite{ou2024mutators}. Across these efforts, a recurring challenge is search collapse: stochastic generation and prompt sensitivity can bias campaigns toward repetitive templates, which reduces exploration unless the design preserves diversity.

\textbf{Evolutionary search and multi-population exploration}. Evolutionary fuzzing \cite{eberlein2020evolutionary} \cite{li2019v} uses feedback signals (often coverage or reachability) to retain and recombine high-value inputs, but single-population search can converge toward dominant structures when feedback is sparse or noisy. Multi-population ``island'' models address this issue by maintaining independent trajectories and periodically exchanging migrants, improving diversity and helping escape local optima.  FunSearch~\cite{romera2024mathematical} and newer works that build on it \cite{novikov2025alphaevolve, lange2025shinkaevolve} provide modern instances of this idea in an LLM setting: multiple islands explore in parallel and exchange high-performing candidates, improving robustness to stochastic generation and local stagnation. While FunSearch targets program synthesis rather than fuzzing, it highlights a practical mechanism that aligns well with LLM-driven input generation, where maintaining multiple distinct trajectories can reduce early collapse to a single coding style.

\textbf{Positioning \tool}. \tool sits at the intersection of LLM-based structured input synthesis and multi-population evolutionary exploration. Compared to Fuzz4All, \tool introduces per-island semantic initialization via distinct seed instructions, which creates multiple topic-level starting points. It then maintains these trajectories through a multi-island evolutionary loop with a migration policy that transfers promising candidates without fully resetting search history, which aims to preserve diversity while still exploiting high-value discoveries. Compared to throughput-oriented compiler fuzzers such as Kitten, \tool trades raw input volume for higher coverage-per-input and complementary failure discovery under model-synthesized programs. Finally, compared to FunSearch, \tool adapts multi-island evolution to a fuzzing context with compiler-specific signals (instrumented compiler coverage and compilation outcomes) and an oracle-driven bug collection pipeline, which grounds the search in the requirements of real compiler testing.

\section{\tool APPROACH}
\begin{figure*}[!htb]
\centering
\includegraphics[width=0.85\textwidth]{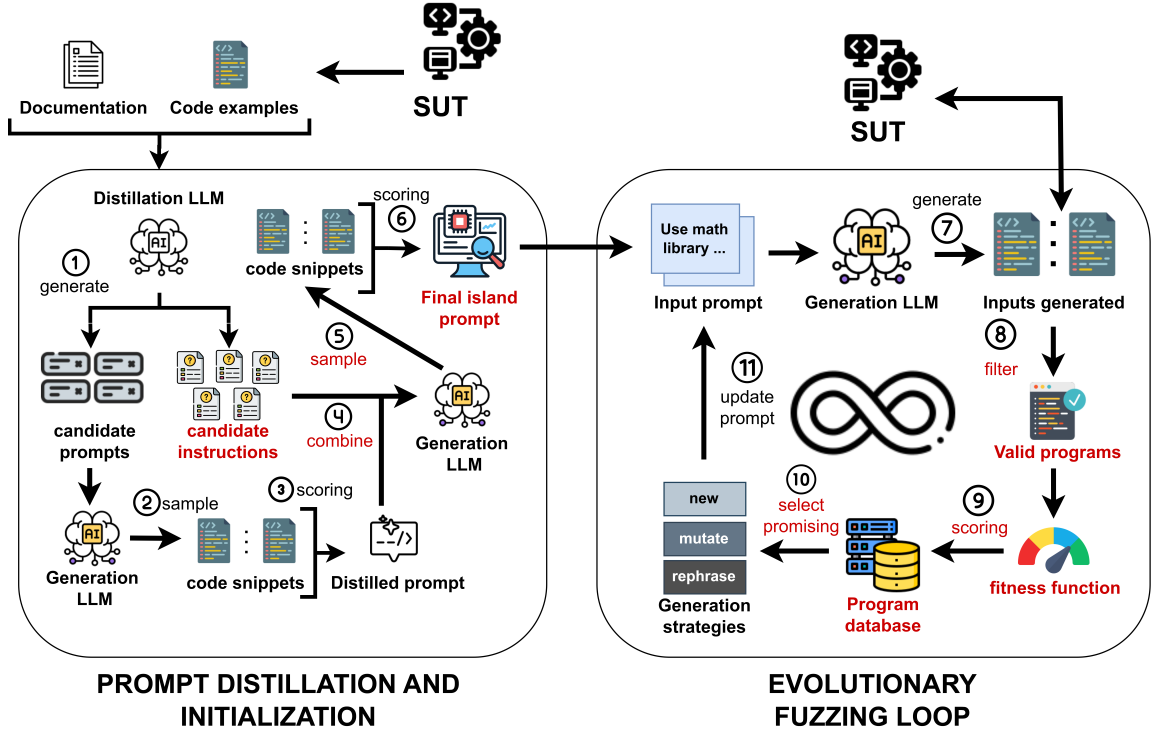}
\caption{\tool strategy: High-level view of the evolutionary loop driven by LLM-generated prompts and feedback-guided scoring. The diagram abstracts away the presence of multiple islands to focus on the core dynamics of a single evolutionary cycle. Red labels denote components introduced in \tool beyond Fuzz4All.}
\label{fig:own_fuzz}
\end{figure*}

%FunFuzz is a two-stage framework that couples documentation-driven LLM program synthesis with a multi-island evolutionary search loop. The goal is to (i) maximize incremental compiler-level coverage during compilation and (ii) collect crash/ICE-triggering inputs; we do not attempt to detect semantic miscompilations. FunFuzz builds on the prompt distillation workflow of Fuzz4All to remain portable across targets (e.g., manuals, specifications, or curated examples) without requiring target-specific grammars. It then adapts the multi-island search principle of FunSearch to the fuzzing setting, where stochastic generation and prompt sensitivity can cause early convergence to repetitive program styles. Figure X summarizes the pipeline.
\tool is a two-stage fuzzing framework that combines documentation-driven LLM input synthesis with a multi-island evolutionary loop. The framework is explicitly designed to be agnostic to the \emph{system under test} (SUT). To this end, we adopt the prompt distillation workflow of Fuzz4All~\cite{xia2024fuzz4all}, which enables the system to ingest heterogeneous sources of target-facing knowledge (manuals, language specifications, or code examples) and transform them into generation prompts. On top of this SUT-agnostic generation layer, we adapt the FunSearch multi-island mechanism~\cite{romera2024mathematical} to fuzzing, allowing multiple partially independent evolutionary processes to explore the SUT in parallel. This design mitigates premature convergence under stochastic LLM generation and prevents early collapse to a single prompt trajectory, while still enabling information sharing across islands. Figure~\ref{fig:own_fuzz} summarizes the overall pipeline.

%Stage 1 (Prompt Distillation and Initialization) converts user-provided artifacts into an initial set of prompts that yield a high rate of compiling programs. A distillation LLM proposes a small set of candidate prompts; a generation LLM samples programs from each candidate; and FunFuzz selects a base prompt using compile-validity (the fraction of samples that compile) as an initialization filter to avoid wasting budget on non-compiling outputs. To prevent the entire campaign from following a single prompt trajectory, FunFuzz also derives island-specific seed instructions that emphasize different topics extracted from the input artifacts, yielding semantically distinct starting points. Stage 2 (Evolutionary Fuzzing Loop) runs parallel fuzzing loops, one per island. Each island repeatedly generates programs from its current prompt, normalizes them deterministically, compiles them against an instrumented compiler, and assigns fitness based on incremental compiler coverage (plus additional signals defined in §X). Parent selection uses a temperature-controlled softmax over fitness to balance early exploration and later exploitation, and selected exemplars are incorporated into subsequent prompts to steer generation. Periodic migration exchanges a limited fraction of high-value candidates from stronger islands to weaker ones while pruning low-value candidates, which mitigates stagnation without fully resetting island state.
During the first stage (\textit{Prompt Distillation and Initialization}), user-provided artifacts are converted into prompts that provide a high rate of compiling programs. A distillation LLM proposes candidate prompts, a generation model samples programs from each candidate, and \tool selects a base prompt using compile-validity as a lightweight initialization signal. To prevent all islands from starting from the same region of the input space, \tool also derives island-specific seed instructions that emphasize different topics extracted from the input documentation. Then, the second stage (\textit{Evolutionary Fuzzing Loop}) runs parallel fuzzing loops across islands. Each island repeatedly generates programs from its current prompt, compiles them against an instrumented compiler, and assigns fitness from incremental compiler coverage plus failure signals. Parent programs are sampled with a temperature-controlled softmax over fitness scores, and the selected parent is incorporated into the next prompt to steer subsequent generations. Islands exchange candidates only during periodic migration events, which transfer high-value programs and prune low-value ones without resetting entire populations.

The following subsections provide a detailed description of the overall \tool approach.%specify the prompt initialization procedure, the per-island loop, the migration policy, and the feedback signals used for fitness and bug detection.

\subsection{Prompt Distillation and Initialization}
\label{sec:autoprompting}
As in Fuzz4All, the pipeline begins by processing user input, which may consist of compiler documentation, protocol specifications, or curated code examples. A \textbf{\textit{Distillation LLM}} summarizes this content into a small, fixed set of \textbf{\textit{candidate prompts}} \circled{1}. Each candidate prompt is evaluated by a \textbf{\textit{Generation LLM}} to synthesize a fixed number of test programs \circled{2}. The validity of each prompt is quantified by computing a compile-validity score, defined as the proportion of generated programs that compile successfully without manual intervention. Compile-validity serves as a lightweight initialization signal; the evolutionary loop later relies on incremental compiler coverage as fitness. The candidate with the highest compile-validity becomes the \textit{\textbf{distilled prompt}} used in subsequent stages~\circled{3}.

In addition to the \textit{\textbf{distilled prompt}}, our system introduces a second prompt layer to promote diversity across islands. Specifically, at step~\circled{1}, we perform an additional LLM call to generate \textbf{\textit{two batches of candidate seed instructions}}, one sampled with low temperature and another with high temperature. 

Each batch contains $N$ instructions, where $N$ corresponds to the number of islands. Each instruction is a short directive that biases program generation toward a specific concept or subsystem (e.g., allocation patterns, control flow, or error-handling constructs). Thus, each batch directly defines a full set of candidate island initializations.

To determine which batch yields better initialization, each instruction is combined with the distilled base prompt~\circled{4}, producing $N$ \textbf{\textit{hybrid prompts}} per batch. These hybrid prompts are passed to the \textbf{\textit{generation LLM}}~\circled{5}, and a fixed budget of programs is synthesized and evaluated for each one. A second round of \textbf{\textit{compile-validity scoring}}~\circled{6} is then performed.

Importantly, selection is performed at the \textbf{batch level}: the scores are aggregated across all hybrid prompts within each batch, and the higher-scoring batch (low-temperature or high-temperature) is selected in its entirety. The selected batch provides one hybrid prompt per island, which becomes the initial prompt for that island.

This two-stage process improves initial compilation throughput while encouraging semantic diversity across islands, reducing the likelihood of premature convergence to homogeneous program structures. Further details are provided in Appendix~\ref{append:autoprompt}.

\subsection{Evolutionary Fuzzing Loop}

\begin{figure*}[!htb]
\centering
\includegraphics[width=0.9\textwidth]{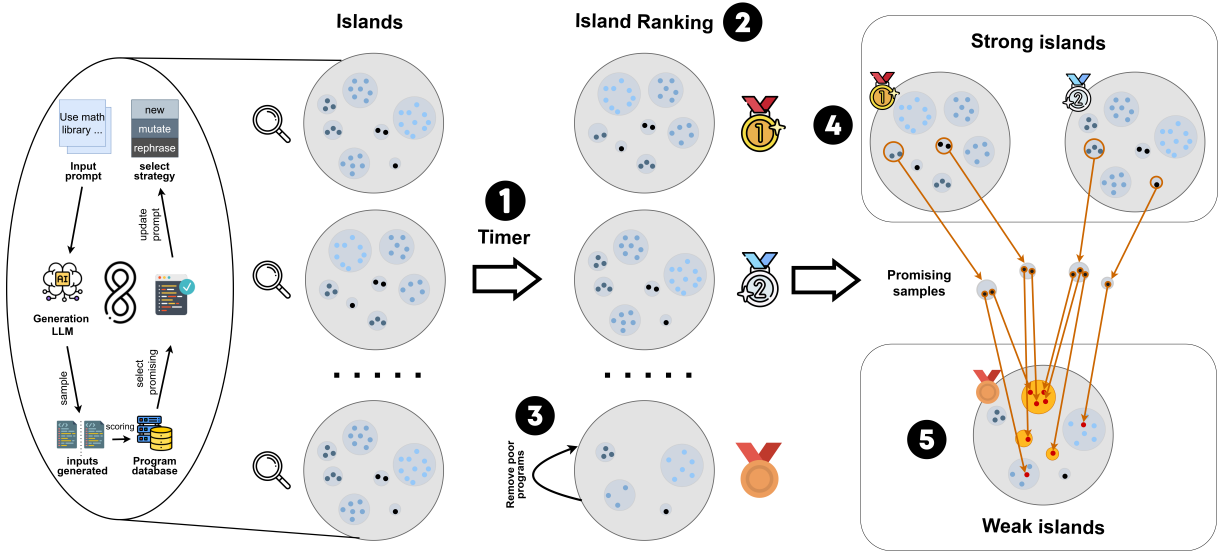} 

\caption{Evolutionary Multi-Island Fuzzing Loop}
\label{fig:islands_behaviours}
\end{figure*}

Inspired by FunSearch, \tool runs multiple evolutionary searches in parallel, one per island (Figure \ref{fig:islands_behaviours}). Each island operates independently, evolving its own population of candidate programs through feedback and guided mutations. %Isolation across islands preserves diversity and reduces the chance that early LLM sampling biases the entire search.
To structure the explanation of this component, we divide it into Island Loop, Cross-Island Migration Policy and Fitness Computation and Scoring Strategy.

For completeness, Appendix~\ref{append:algorithm} provides a high-level algorithmic summary of the worker-driven fuzzing loop, making explicit how LLM-driven generation, island-local evaluation, migration, and SUT-level failure detection are orchestrated during execution.

\subsubsection{Island Loop}
\label{sec:islandloop}
An \textit{island} in our system represents an independent evolutionary unit that maintains its own population of candidate programs. Each program in the population is assigned a fitness score based on compiler feedback, as described in Section~\ref{sec:fitness}. %To maintain compatibility with the underlying FunSearch formulation, programs within an island are grouped into score-based clusters.

Figure~\ref{fig:own_fuzz} shows the lifecycle executed by each island. The loop starts with the \textbf{\textit{final island prompt}} produced in the previous stage, so that it represents the initial \textbf{\textit{input prompt}} in this stage. The \textbf{\textit{generation LLM}} samples a batch of candidate programs \circled{7}. Before compilation, \tool applies a deterministic \emph{normalization} to remove common LLM formatting artifacts and to standardize compilation units. Specifically, the normalization (i) drops malformed or non-resolvable \texttt{\#include}, (ii) adds standard-library includes only when the compiler reports missing declarations, using a fixed mapping from diagnostic patterns to headers, and (iii) splits multi-program outputs into separate compilation units. This normalization does not perform semantic repair; it only makes compilation outcomes comparable across candidates.

\tool compiles each normalized candidate and applies a lightweight failure oracle \circled{8}. The oracle flags abnormal compiler termination using exit status and compiler diagnostics. Successful compilation and standard compilation errors are treated as benign outcomes; other exit codes or internal-failure indicators in stderr (e.g., ``internal compiler error'') are treated as potential bugs. The oracle targets crashes and internal errors rather than semantic miscompilations.

The island assigns each candidate a fitness score \circled{9} (Section~\ref{sec:fitness}). \tool selects a promising candidate by using a softmax distribution over fitness \circled{10}, with a temperature schedule that shifts from exploration to exploitation over time. This probabilistic mechanism biases sampling toward higher-scoring programs while preserving variability, particularly during early exploration phases. As the temperature decays, sampling gradually shifts toward exploitation, allowing each island to refine its search trajectory.

The next prompt incorporates (i) the distilled base prompt, (ii) one selected program, and (iii) a transformation instruction that controls the generation mode (new, mutate, rephrase) \circled{11}. \tool updates the prompt and repeats the loop. To avoid prompt overfitting to a small set of programs, the system penalizes or removes programs after they appear as prompt programs to encourage the system to explore unexplored behaviors.

%The multi-island design enables parallel exploration under a fixed wall-clock budget. Section~\ref{sec:fairness} reports throughput and discusses when additional islands yield measurable speedups.

\subsubsection{Cross-Island Migration Policy}

To promote diversity and avoid premature convergence, \tool introduces a periodic migration mechanism across islands by extending the design of FunSearch. As depicted in Figure~\ref{fig:islands_behaviours}, this migration step is triggered every 3 hours~\circledblack{1}. We rank islands by coverage and designate the top 51\% (rounded up) as strong; the remaining islands form the weak set. Unlike FunSearch, which resets weak islands, \tool applies a soft recovery step: in each weak island, we remove the lowest-scoring 30\% of the population (according to the island's fitness scores), preserving the remaining individuals and their accumulated search context.

%Unlike FunSearch, where weak islands are fully reset, our approach applies a softer recovery strategy: only the bottom 30\% of programs~\circledblack{3} in each weak island are removed. This selective pruning preserves part of the island’s search history and structural patterns—often still useful—while discarding outdated programs that are unlikely to contribute further.

Next, each strong island~\circledblack{4} shares 10\% of its population. Migrants are sampled from the island's elite pool (the top 20\% by fitness), which biases transfer toward high-value programs while maintaining diversity among the exported set. Migrants are injected into weak islands to replace the pruned individuals. Then, the weak islands are populated with the shared programs~\circledblack{5}, which are re-evaluated using the local fitness criteria. Since each island maintains its own scoring state, transferred candidates are re-evaluated using the local coverage and statistics of the receiving island. This discrepancy helps challenge the search boundaries of weaker islands and may uncover new directions previously overlooked.

We use a fixed migration schedule and fixed transfer/pruning ratios across all experiments. Unless otherwise stated, migrations occur every 3 hours; weak islands are partially pruned and then replenished with a small sample of high-scoring programs from strong islands. This “soft migration” is intended to mitigate stagnation without discarding island state.

These hyperparameters (migration period, pruning ratio, sharing rate, and number of islands) were selected empirically based on preliminary experiments and kept fixed across all evaluations to ensure a fair and controlled comparison.

We intentionally do not perform an exhaustive sensitivity analysis, as our primary focus is on evaluating the overall effectiveness of the proposed approach rather than fine-tuning individual parameters. While this choice simplifies the experimental design, the selected values may not generalize across different targets, workloads, or hardware configurations. A detailed list of all hyperparameters is provided in Appendix \ref{append:experiments_configuration}.

\subsubsection{Fitness Computation and Scoring Strategy}
\label{sec:fitness}
%The primary fitness signal used across all islands is the incremental compiler coverage triggered when compiling a generated program, which serves as a proxy for behavioral novelty. Higher coverage implies a broader exploration of the target's execution paths and therefore a more valuable candidate for retention and recombination.
Our primary fitness signal is \textit{marginal compiler coverage}: for each generated program, we compile it with an instrumented build of the compiler-under-test and measure the additional compiler source lines exercised by this compilation relative to the island's current coverage set. This marginal gain serves as a proxy for behavioral novelty and is used to prioritize candidates for retention and reuse.

Coverage is collected at the compiler level (the compiler is the SUT), and each island maintains its own local coverage frontier. Using per-island frontiers encourages partially independent exploration: islands can pursue different behavioral regions without being immediately dominated by discoveries made by other islands. Programs that yield positive marginal gains are preferentially retained and sampled as parents, while repeatedly non-contributing programs are deprioritized. We also evaluate optional scoring modifiers (e.g., failure handling, compilation-time rewarding, redundancy filtering) in Appendix~\ref{append:fitness} and in the ablation study.

\section{Experimental Evaluation}
\label{sec:experiments}
This section evaluates \tool and analyzes the impact of its core design choices. Our experimental campaign addresses four questions: (1) how much compiler coverage \tool achieves relative to state-of-the-art fuzzers, (2) how effectively it discovers actionable compiler failures over long runs, (3) which components drive its performance through controlled ablations, and (4) whether it can be steered toward specific language features without collapsing exploration.

\textbf{Hardware and Execution Environment.}
All fuzzing frameworks are executed inside Docker containers on the same host machine (MSI Vector GP77 13V; Intel i7-13700H; NVIDIA RTX 4060). For LLM-based fuzzers (\tool and Fuzz4All), model inference is served from a dedicated GPU server (NVIDIA A100 80GB), while compilation, coverage measurement, and crash triage are executed locally on the host machine. Kitten is executed entirely on the local host.

\textbf{Input Pre-processing and Normalization.}
To ensure comparability between \tool and Fuzz4All, we apply the same deterministic normalization step to all LLM-generated outputs prior to compilation (Section \ref{sec:islandloop}). This step only standardizes compilation units (e.g., includes and splitting) and is applied uniformly.

\textbf{Coverage Instrumentation.}
We measure compiler-level coverage by compiling each generated program with an instrumented build of the target compiler and recording cumulative unique source-line coverage within the compiler codebase itself, rather than coverage of the generated test programs. All fuzzers are evaluated against the same instrumented compiler binaries and the same coverage collection pipeline. For GCC we use an instrumented build of the gcc-16.0.0 version\footnote{https://github.com/gcc-mirror/gcc commit 94e21bcec061c1691cd0d5382e3c6f420e0386c4}; for Clang/LLVM we use an instrumented build of a pinned trunk snapshot\footnote{LLVM commit 94e21bcec061c1691cd0d5382e3c6f420e0386c4} to keep results reproducible. We report cumulative covered source lines aggregated across all compilations in a run.

\textbf{Execution Policy.}
Each experimental configuration is executed three times with different random seeds. Reported results correspond to averages across these runs, mitigating stochastic effects introduced by evolutionary selection and, where applicable, LLM sampling.

\textbf{LLMs Used.}
\tool and Fuzz4All use the same generation LLM and decoding parameters during fuzzing (DeepSeek-Coder-V2-Lite-Base; temperature 1.0; top-p sampling; max 512 tokens), served via vLLM. Prompt distillation uses GPT-4.1 for instruction construction only (not for program sampling during fuzzing).

\textbf{Experiment Budgets.}
For the main evaluation, we run long campaigns of 24 hours to assess coverage and bug-finding performance. The 24-hour budget follows standard fuzzing practice to capture steady-state behavior and enable fair comparison \cite{xia2024fuzz4all} \cite{xie2025kitten}.

For controlled design analysis and ablation studies, we use a budget of 30,000 generated programs. Compared to the 10,000 programs used in Fuzz4All, this increase reflects the multi-island design of our method, which requires sufficient samples per island to properly exploit semantic diversity. The budget remains within the same order of magnitude as prior work, ensuring comparability while avoiding under-sampling.

\textbf{Fitness function.}
The fitness function used across all long-running experiments is described in Section~\ref{sec:scoring_ablation}.

\subsection{Systems Under Test and Baselines}
\label{sec:suts_baselines}

Our evaluation targets modern, widely deployed compiler toolchains whose complexity and continuous evolution make them representative fuzzing subjects.

\textbf{Systems Under Test.}
We evaluate GCC and Clang/LLVM, two mature and actively maintained compilers with deep compilation pipelines (front-end parsing/semantic analysis, IR transformations, and back-end code generation). When we report whether a bug is “not fixed in trunk”, we reproduce crashes on a pinned trunk snapshot to keep results reproducible.

\textbf{Baselines.}
We compare \tool against two state-of-the-art compiler fuzzers: (i) \textbf{Fuzz4All}, an LLM-driven evolutionary fuzzer, which we evaluate in both single-instance and \textit{parallel} (5 instances) configurations; and (ii) \textbf{Kitten}, a high-throughput mutational fuzzer that utilizes a curated seed corpus (sourced from open-access LLVM test suites). To isolate the impact of this external initialization, we evaluate \tool in two distinct modes: a default \textbf{from-scratch} configuration and a \textbf{warm-start} (or \textit{seeded}) configuration, which leverages the same publicly available corpus to initialize its evolutionary loop (further explained in Appendix \ref{appendix:hybridseed}). This allows us to distinguish between gains from \tool's search strategy and gains from target-specific prior knowledge.

\begin{table}[htb!]
\centering
\small
\resizebox{\linewidth}{!}{
\begin{tabular}{lcccccc}
\toprule
\textbf{Approach} & \textbf{LLM-guided} & \textbf{Evolutionary} & \textbf{Curated corpus} & \textbf{Domain init.}  \\
 & \textbf{generation} & \textbf{scoring} & \textbf{required} & \textbf{required}  \\
\midrule
Fuzz4All & \checkmark & $\times$   & $\times$ & \checkmark  \\
Fuzz4All parallel & \checkmark & $\times$ & $\times$ & \checkmark  \\
Kitten & $\times$ & heuristic-based  & \checkmark & \checkmark  \\
FUNFUZZ & \checkmark & \checkmark  & $\times$ & \checkmark  \\
FUNFUZZ (warm-start) & \checkmark & \checkmark  & \checkmark & \checkmark  \\
\bottomrule
\end{tabular}
}
\caption{Comparison of design characteristics.}
\label{tab:design_comparison}
\end{table}

\textbf{Scope of Comparison.}
Not all baselines support all targets in our setting. Kitten targets C and does not provide a C++ fuzzing workflow comparable to our evaluation setup; therefore, C++ experiments compare only \tool and Fuzz4All, while C experiments include all three fuzzers. This selection enables a balanced comparison between LLM-driven and high-throughput mutational fuzzing under realistic constraints. We do not include additional generator-based compiler fuzzers (e.g., Csmith/GrayC for C and YARPGen for C++) in our main comparison; it should be noted that previous work already evaluates Fuzz4All against these baselines in 24-hour campaigns \cite{xia2024fuzz4all}, and Kitten motivates high-throughput mutation as a more comparable baseline when assessing LLM-driven compiler fuzzing; we therefore focus on one strong representative per paradigm. 

\subsection{Fairness and Throughput Considerations}
\label{sec:fairness}

Comparing compiler fuzzers fairly is challenging because LLM-driven fuzzers and high-throughput mutational fuzzers differ substantially in resource constraints and input generation throughput. We therefore report results under two complementary regimes.

\textbf{Throughput reporting and input budgets.}
Rather than throttling faster systems or inflating slower ones, we evaluate each fuzzer under its intended operating mode and explicitly report both wall-clock time and the number of compiler invocations (post-normalization) executed in each experiment. This makes throughput differences transparent and enables input-budget-controlled comparisons where needed.

\textbf{Interpretation of Baselines.}
The baselines considered in this work rely on different assumptions regarding input generation and prior knowledge. In particular, Kitten should be interpreted as a strong mutation-based baseline that benefits from a curated, compiler-specific seed corpus, rather than as a from-scratch generator. This initialization provides a substantial prior over valid and diverse program structures, which directly contributes to its high throughput and early coverage gains. In contrast, \tool in its main configuration does not rely on curated corpora and instead constructs inputs from scratch using LLM-guided generation and evolutionary selection. For this reason, comparisons should be interpreted not only in terms of raw throughput, but also in terms of coverage efficiency and dependence on target-specific initialization. The warm-start configuration further illustrates that curated corpora and LLM-guided evolutionary search are complementary, rather than mutually exclusive.

\textbf{Cost-Benefit Perspective.}
The benefit of \tool should not be interpreted solely through raw generation throughput. While LLM-guided fuzzing incurs additional computational cost due to model inference, its objective is not merely to maximize the number of generated programs, but to improve the effectiveness of each evaluated input and reduce dependence on handcrafted, target-specific initialization. This is why we report both time-based and input-budget-controlled comparisons. Under equalized input budgets, \tool consistently achieves higher compiler coverage than Fuzz4All, indicating better coverage yield per evaluated program. Relative to Kitten, the main advantage of \tool is not raw throughput, but the ability to operate from scratch without requiring a curated compiler-specific corpus, while still remaining competitive in long-running campaigns. Moreover, the warm-start results show that when such prior information is available, \tool can exploit it effectively and achieve substantially higher coverage, suggesting that LLM-guided evolutionary search complements curated initialization rather than replacing it.

\textbf{Configuration Scope.}
For the experiments, we report (i) a single-LLM configuration aligned with Fuzz4All, and (ii) a higher-throughput configuration that increases generation parallelism to approach the input rate of mutational fuzzers. This separation allows us to distinguish gains due to exploration strategy from gains due to input volume.

\subsection{Coverage Evaluation Over 24 Hours}
\label{sec:coverage_results}

We evaluate the coverage achieved by \tool over 24-hour fuzzing campaigns and compare it against the selected baselines under the configurations described in Section~\ref{sec:fairness}. All results are averaged over three independent runs.

\subsubsection{C Coverage Results}
\label{sec:c_coverage}

\begin{figure}[htb!]
    \centering
    % Gráfico de GCC arriba
    \begin{subfigure}[b]{0.95\linewidth}
        \centering
        \includegraphics[width=\linewidth]{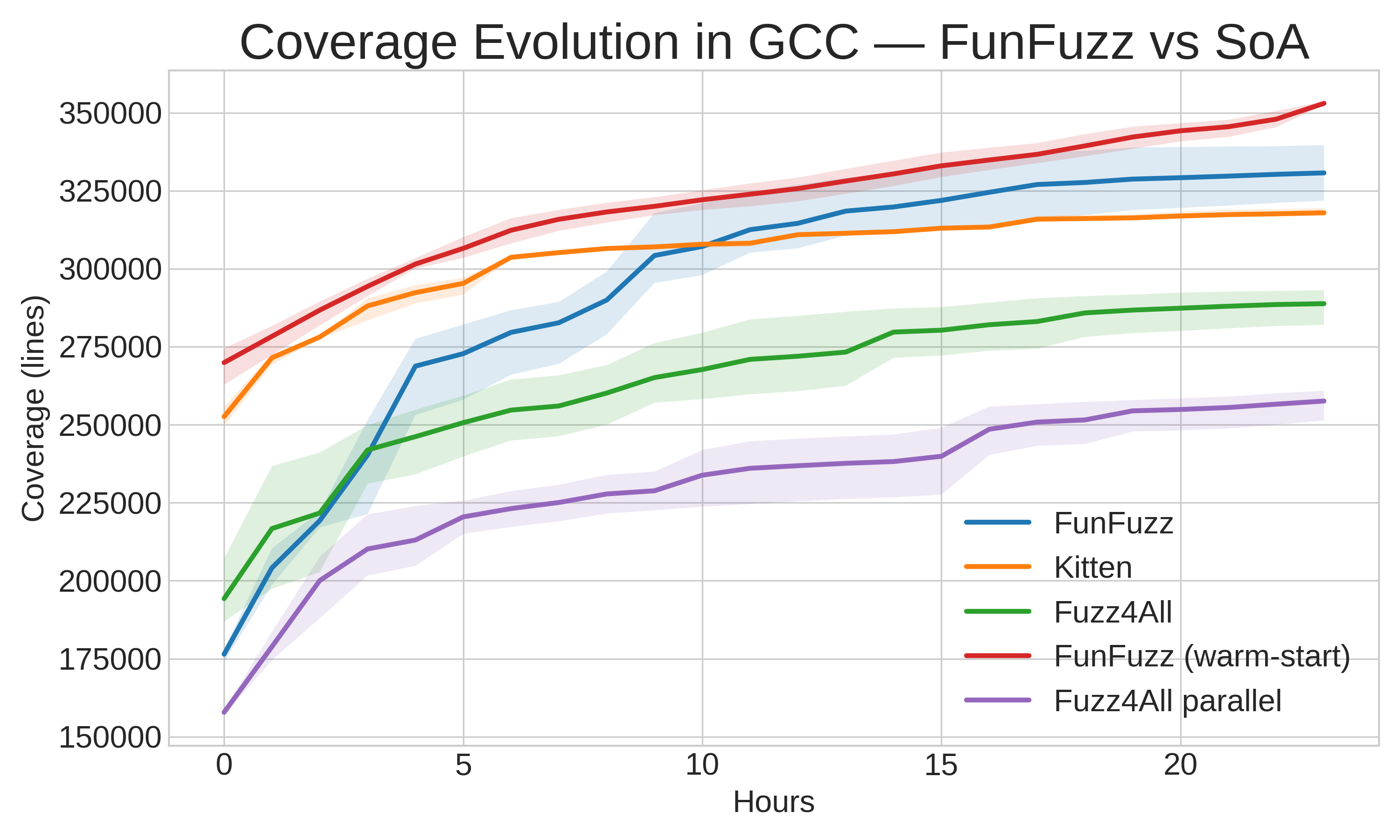} % Cambia al de GCC
        \caption{GNU Compilers (GCC)}
        \label{fig:coverage_gcc}
    \end{subfigure}
    
    \vspace{0.1cm} % Separación vertical para que no se peguen las letras
    
    % Gráfico de Clang abajo
    \begin{subfigure}[b]{0.95\linewidth}
        \centering
        % Cambia este nombre al del archivo de Clang
        \includegraphics[width=\linewidth]{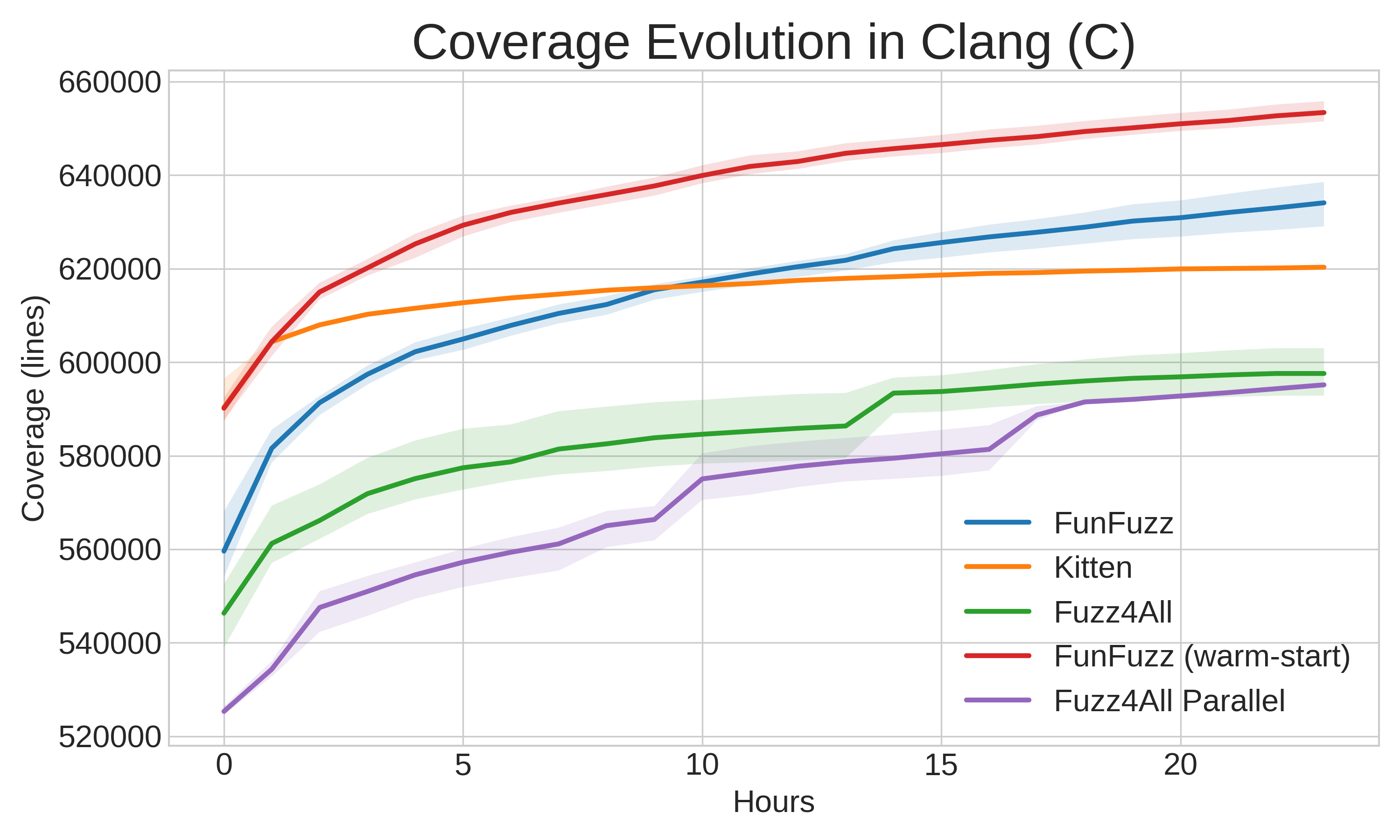} 
        \caption{LLVM Compilers (Clang)}
        \label{fig:coverage_clang}
    \end{subfigure}
    
    % El caption global va fuera de los subfigures
    \caption{Coverage evolution over 24 hours in \textbf{C}. \tool significantly outperforms Fuzz4All and keeps improving beyond Kitten in both environments.}
    \label{fig:coverage_c}
\end{figure}

Figure~\ref{fig:coverage_c} plots 24-hour coverage trajectories for both GCC-16.0.0 and LLVM/Clang (C) infrastructures. Across both compiler environments, \tool overtakes Fuzz4All early in the campaign and continues to increase steadily throughout the run, with no clear saturation trend. Kitten improves quickly at the beginning but plateaus earlier. The warm-start variant (\tool (warm-start)) achieves the highest curve across both targets by leveraging Kitten’s curated corpus. Shaded regions indicate variability across three runs.

Furthermore, Table~\ref{tab:c_results_combined} summarizes final coverage across both compilers under two evaluation settings: (i) program-budget-aligned (same number of compiler invocations post-normalization) and (ii) 24-hour time-based comparison. Under the same input budget of 238,222 invocations, \tool (5 islands) reaches 328,188 and 632,670 covered lines for GCC and Clang, versus 292,872 and 597,822 for Fuzz4All (+12.1\% and +5.8\% improvements, respectively). The warm-start variant further elevates these budget-aligned maximums to 350,519 and 641,707 lines (+19.7\% and +7.3\%). This demonstrates a higher coverage discovery per input independent of throughput. This advantage emerges early, and \tool reaches coverage levels comparable to Fuzz4All’s end-of-run results substantially earlier in the campaign. Over 24 hours, \tool achieves 329,905 (GCC) and 634,091 (Clang) covered lines, exceeding both standard Fuzz4All and Kitten (316,900 in GCC, 620,312 in Clang), despite Kitten executing substantially more inputs (857,567 invocations). This difference is consistent with Kitten operating over a curated, compiler-specific seed corpus, which reduces the need for initial exploration and allows it to focus on exploiting existing program structures. However, this advantage comes with a strong dependence on the availability of high-quality curated seeds, whereas \tool is able to generate programs from scratch. While LLM-driven generation introduces run-to-run variance, \tool consistently dominates across seeds and compilation targets, suggesting that the evolutionary loop is robust to imperfect early generations.

More importantly, to establish a fair multi-process baseline, we evaluate a \textit{Fuzz4All parallel} configuration which launches 5 isolated instances concurrently to mimic our 5-island hardware budget. Surprisingly, Fuzz4All parallel performs substantially worse than a single standard Fuzz4All instance on both platforms, dropping to 257,593 lines in GCC and 595,161 in Clang. This degradation suggests that simply scaling the number of independent instances does not translate into better exploration; rather, it induces significant redundancy, with many generated test cases effectively exploring the same regions of the input spaces. In contrast, our evolutionary multi-island approach explicitly prevents this issue: through periodic migration and fitness-based pruning, \tool gracefully distributes the search effort and effectively shares independent discoveries without suffering from massive overlap. Consequently, multi-island execution contributes substantially to our approach, with 5 cooperative islands heavily outperforming a single isolated island under the same 24-hour budget (329,905 vs. 303,193 in GCC, and 634,091 vs. 623,560 in Clang). Finally, \tool warm-start significantly boosts performance across both evaluation regimes, achieving the global maximums of 372,275 and 653,402 covered lines for GCC and Clang, respectively.

\begin{table}[htb!]
\centering
\setlength{\tabcolsep}{2pt}
\renewcommand{\arraystretch}{1.08}
% Adapta el tamaño final para que encaje 100% perfecto en una sola columna:
\resizebox{\linewidth}{!}{
\begin{tabular}{l|r|r|r|r}
\toprule
& & & \multicolumn{2}{c}{\textbf{Coverage}} \\
\cmidrule(lr){4-5}
\textbf{Configuration} & \textbf{Progs.} & \textbf{Valid \%} & \textbf{GCC} & \textbf{Clang} \\
\midrule
\multicolumn{5}{c}{\textbf{Program-budget-aligned comparison}} \\
\midrule
Fuzz4All             & 238\,222 & 42.60 & 292\,872 & 597\,822 \\
\tool (5 islands)    & 238\,222 & 47.28 & \textbf{328\,188} {\color{darkgreen}\scriptsize(+12.1\%)} & \textbf{632\,670} {\color{darkgreen}\scriptsize(+5.8\%)} \\
\tool (warm-start) & 238\,222 & 43.48 & \textbf{350\,519} {\color{darkgreen}\scriptsize(+19.7\%)} & \textbf{641\,707} {\color{darkgreen}\scriptsize(+7.3\%)} \\
\midrule
\multicolumn{5}{c}{\textbf{Time-based comparison (24 hours)}} \\
\midrule
Fuzz4All             & 238\,222 & 42.60 & 292\,872 & 597\,822 \\
Fuzz4All parallel    & 249\,456 & 41.60 & 257\,593 & 595\,161 \\
Kitten               & 857\,567 & 17.16 & 316\,900 & 620\,312 \\
\tool (1 island)     & 240\,257 & 47.90 & \textbf{303\,193} {\color{darkgreen}\scriptsize(+3.5\%)} & \textbf{623\,560} {\color{darkgreen}\scriptsize(+4.3\%)} \\
\tool (5 islands)    & 252\,283 & 47.30 & \textbf{329\,905} {\color{darkgreen}\scriptsize(+12.7\%)} & \textbf{634\,091} {\color{darkgreen}\scriptsize(+6.1\%)} \\
\tool (warm-start) & 479\,888 & 43.03 & \textbf{372\,275} {\color{darkgreen}\scriptsize(+27.1\%)} & \textbf{653\,402} {\color{darkgreen}\scriptsize(+9.3\%)} \\
\bottomrule
\end{tabular}
}
\caption{Coverage and throughput results for \textbf{C}. Comparisons isolate the effect of program-budget constraints and strictly 24-hour scenarios evaluating both GCC and LLVM infrastructures. Percentage improvements are reported relative to Fuzz4All within the same compilation target.}
\label{tab:c_results_combined}
\end{table}

\subsubsection{C++ Coverage Results}
\label{sec:cpp_coverage}

\begin{figure}[htb!]
    \centering
    % Gráfico de GCC arriba
    \begin{subfigure}[b]{0.95\linewidth}
        \centering
        \includegraphics[width=\linewidth]{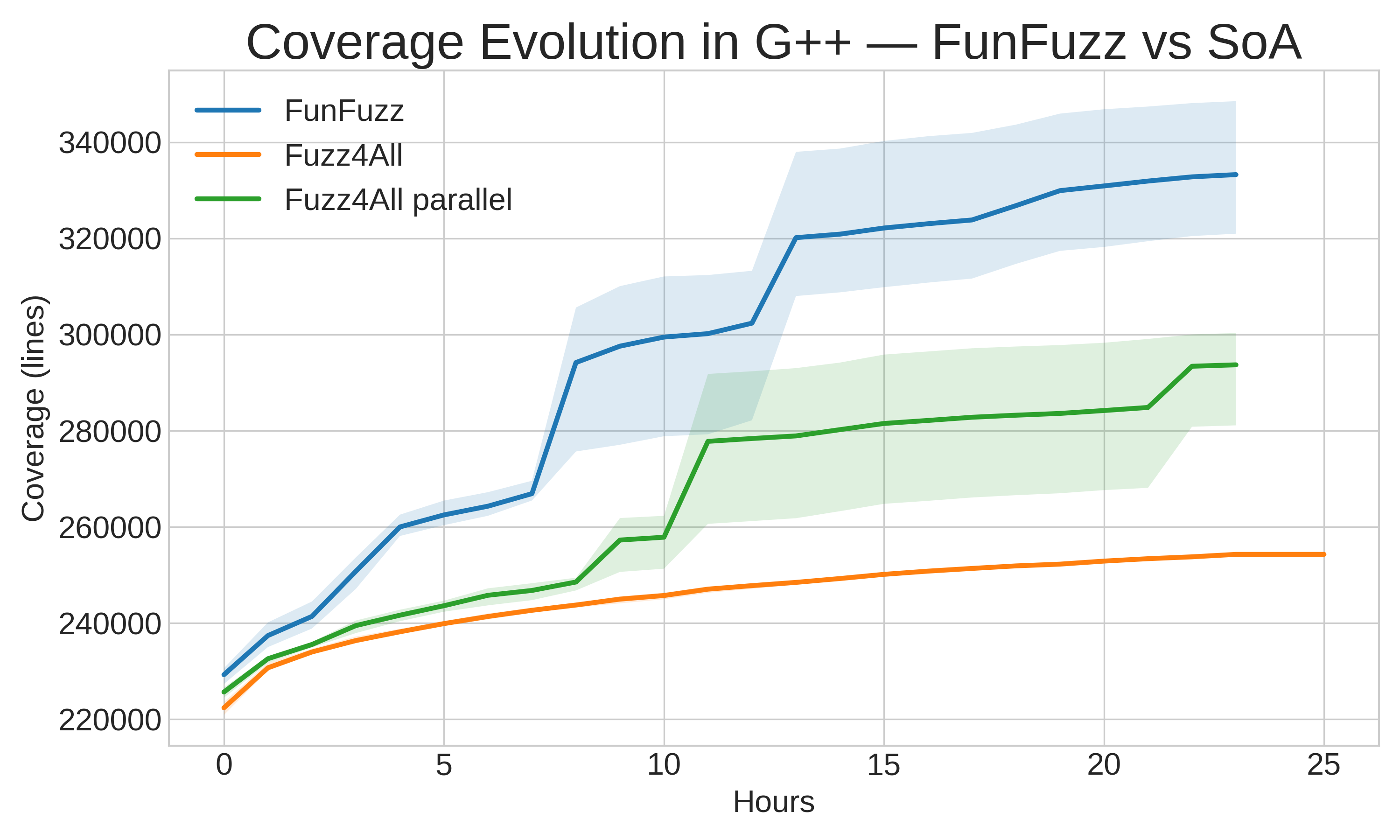}
        \caption{GNU Compilers (G++)}
        \label{fig:coverage_gcc_cpp}
    \end{subfigure}
    
    \vspace{0.1cm} % Separación vertical
    
    % Gráfico de Clang abajo
    \begin{subfigure}[b]{0.95\linewidth}
        \centering
        \includegraphics[width=\linewidth]{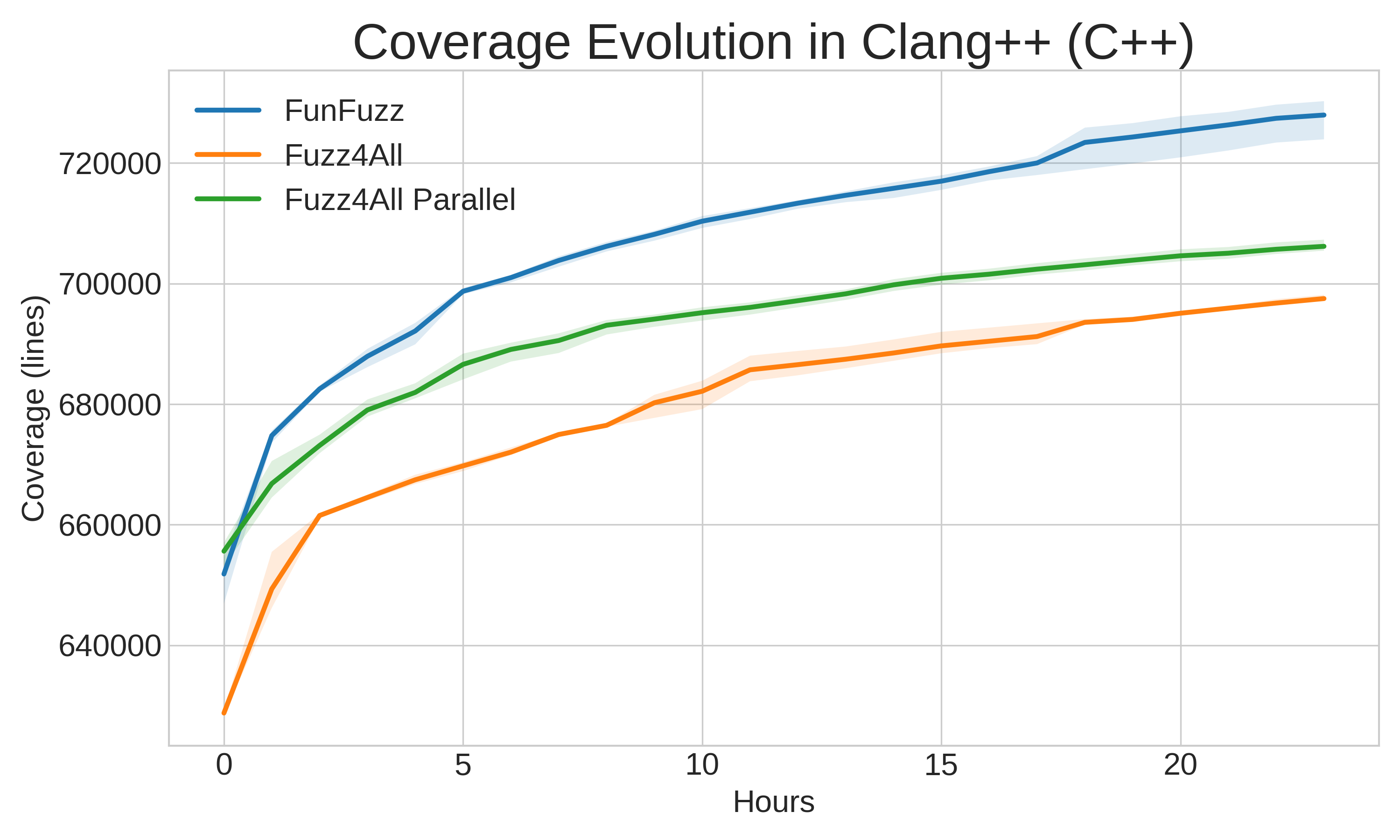} 
        \caption{LLVM Compilers (Clang++)}
        \label{fig:coverage_clang_cpp}
    \end{subfigure}
    
    \caption{Coverage evolution over 24 hours in \textbf{C++}. \tool shows a clear and sustained advantage over Fuzz4All and its parallel counterpart in both environments.}
    \label{fig:coverage_cpp}
\end{figure}

Figure~\ref{fig:coverage_cpp} plots 24-hour coverage trajectories for both G++ and LLVM/Clang (C++) infrastructures. Across both compiler environments, \tool establishes an early lead over Fuzz4All, and the gap persists throughout the full campaign, indicating sustained gains beyond initial exploration effects (shaded regions denote variability across valid runs). Table~\ref{tab:cpp_results_combined} reports final coverage under two settings: (i) an input-budget-controlled comparison and (ii) a 24-hour time-based comparison. Under the same budget of 91,561 invocations, \tool (5 islands) reaches 290,199 covered lines versus 254,247 for Fuzz4All (+14.1\%) in G++, and 707,638 versus 697,519 (+1.5\%) in Clang++, demonstrating higher coverage discovery per input independent of throughput.

Over 24 hours, \tool achieves 333,254 covered lines in G++ (+31.0\%) and 727,964 in Clang++ (+4.4\%). Notably, even a single evolutionary island substantially outperforms the standard baseline in both environments (reaching 323,145 covered lines in G++; +27.1\% and 722,324 in Clang++; +3.6\%), confirming that evolutionary scoring alone provides strong benefits, while multi-island parallelism adds further gains. Furthermore, we evaluate the \textit{Fuzz4All parallel} multi-process baseline. In contrast to C (where redundant overlap caused negative scaling), launching independent Fuzz4All instances in C++ yields some improvement over a single instance (reaching 293,708 lines in G++ and 706,185 in Clang++). Nonetheless, \tool (5 islands) robustly outperforms this naive parallelization approach across both targets. These improvements are particularly notable for C++, where templates and richer language semantics expose deeper and more structurally complex compilation paths, confirming that \tool's collaborative search effectively avoids saturation and generalizes well across languages.

\begin{table}[htbp]
\centering

\setlength{\tabcolsep}{2pt}
\renewcommand{\arraystretch}{1.08}
\resizebox{\linewidth}{!}{
\begin{tabular}{l|r|r|r|r}
\toprule
& & & \multicolumn{2}{c}{\textbf{Coverage}} \\
\cmidrule(lr){4-5}
\textbf{Configuration} & \textbf{Progs.} & \textbf{Valid \%} & \textbf{G++} & \textbf{Clang++} \\
\midrule
\multicolumn{5}{c}{\textbf{Program-budget-aligned comparison}} \\
\midrule
Fuzz4All             & 91\,561  & 46.42 & 254\,247 & 697\,519 \\
\tool (5 islands)    & 91\,561  & 38.20 & \textbf{290\,199} {\color{darkgreen}\scriptsize(+14.1\%)} & \textbf{707\,638} {\color{darkgreen}\scriptsize(+1.5\%)} \\
\midrule
\multicolumn{5}{c}{\textbf{Time-based comparison (24 hours)}} \\
\midrule
Fuzz4All             & 91\,561  & 44.46 & 254\,298 & 697\,519 \\
Fuzz4All parallel    & 195\,162 & 38.80 & 293\,708 & 706\,185 \\
\tool (1 island)     & 205\,874 & 46.53 & \textbf{323\,145} {\color{darkgreen}\scriptsize(+27.1\%)} & \textbf{722\,324} {\color{darkgreen}\scriptsize(+3.6\%)} \\
\tool (5 islands)    & 225\,854 & 44.46 & \textbf{333\,254} {\color{darkgreen}\scriptsize(+31.0\%)} & \textbf{727\,964} {\color{darkgreen}\scriptsize(+4.4\%)} \\
\bottomrule
\end{tabular}
}
\caption{Coverage and throughput results for \textbf{C++}. Comparisons evaluate both G++ and Clang++ infrastructures. Percentage improvements are reported relative to standard Fuzz4All within the same target compiler.}
\label{tab:cpp_results_combined}
\end{table}

\subsection{Bug-Finding Effectiveness}
\label{sec:bug_finding}

While coverage serves as a proxy for exploration, the ultimate goal of a compiler fuzzer is to discover actionable bugs. We therefore evaluate bug-finding effectiveness under long-running campaigns and report results over three independent 24-hour executions per configuration. Bugs are counted after post-processing and deduplication, as described below. When we refer to the GCC trunk version, we mean the latest development version corresponding to the current HEAD commit of the official GCC GitHub mirror, specifically commit d78b2b6c01243c59fc52937e1e3b0d84848a8fa9 at the time of our experiments.
\subsubsection{Crash Deduplication and Unique Bug Identification}
\label{sec:crash_dedup}

During fuzzing, many inputs may trigger the same underlying defect with minor syntactic variations. To avoid overcounting and enable consistent comparison across fuzzers, we apply a crash deduplication pipeline. Each candidate input is recompiled under a fixed timeout and configuration using the target compiler, and we retain executions that exhibit \emph{compiler-internal failures} (e.g., ICE-style diagnostics, assertion failures, or crash-like abnormal termination), rather than semantic miscompilations.

For each internal failure, we extract (i) the primary diagnostic header describing the failure and (ii) an associated stack trace or backtrace when available. Both are normalized by removing nondeterministic tokens (such as file paths, line numbers, and memory addresses). We report \textit{Unique} as the number of distinct fingerprints produced by this procedure. 

\subsubsection{Bug Budget Search on C Compilers}
\label{sec:bug_budget_gcc}
To assess bug-finding capability beyond coverage, we run a bug-budget evaluation on the C frontends of GCC and Clang, using gcc-16.0 and clang-23, respectively. For each configuration, we perform three independent 24-hour runs under identical settings.

For GCC (a released version), we additionally report Not fixed in Trunk, i.e., the subset of unique failures that still reproduce on a pinned trunk snapshot used for reproducibility.

\begin{table}[htb!]
\centering
%\small
\setlength{\tabcolsep}{3.5pt}
\renewcommand{\arraystretch}{1.08}
\begin{tabular}{C{1.5cm}|C{1.2cm}|C{1cm}|C{1.4cm}|C{1.4cm}}
\toprule
\textbf{} & \textbf{Fuzz4All} & \textbf{Kitten} & \textbf{\tool} & \textbf{\tool (warm-start)} \\
\midrule
\multicolumn{5}{c}{\textbf{GCC-16.0 (C frontend)}} \\
\midrule
Unique & 5  & 9  & 14 & 34 \\
Not fixed in Trunk & 3 & 8 & 9 & 28 \\
\midrule
\multicolumn{5}{c}{\textbf{Clang-23 (C frontend)}} \\
\midrule
Unique & 14 & 26 & 24 & 52 \\
\bottomrule
\end{tabular}
\caption{Unique compiler-internal failures over three independent 24-hour runs on C compilers. ``Not fixed in Trunk'' is reported only for GCC.}
\label{tab:bugs_c_compilers}
\end{table}

\textbf{GCC}. Table~\ref{tab:bugs_c_compilers} shows that \tool uncovers substantially more unique GCC failures than both baselines: 14 unique bugs versus 9 for Kitten and 5 for Fuzz4All. The same trend holds when checking persistence on trunk: \tool reproduces 9 unique failures on trunk, compared to 8 for Kitten and 3 for Fuzz4All. The warm-start configuration yields a much larger number of unique and trunk-reproducible failures (34 / 28), illustrating that \tool can effectively exploit richer initialization; however, this setting is not baseline-aligned and should be interpreted as a compatibility/upper-bound variant rather than a direct comparison.

\textbf{Clang}. Table~\ref{tab:bugs_c_compilers} reports the corresponding results for clang-23. Because Clang experiments are conducted on trunk snapshots, we report only the number of unique deduplicated failures. \tool again significantly improves over the LLM baseline (24 unique vs. 14 for Fuzz4All), while Kitten remains highly competitive on this target (26 unique). As for GCC, warm-start configuration substantially increases the number of discovered unique failures (52), suggesting that \tool's evolutionary loop scales with stronger or more diverse seed corpora, though these results are not baseline-aligned.

\subsubsection{Bug Budget Search on C++ Compilers}
\label{sec:bug_cpp}
We next evaluate bug-finding on the C++ frontends of GCC and Clang (\texttt{g++} and \texttt{clang++-23}) under the same bug-budget protocol as for C, reporting \textbf{Unique} compiler-internal failures after deduplication (Section~\ref{sec:crash_dedup}); for \texttt{g++} we also report \textbf{Not fixed in Trunk}. Kitten does not provide a C++ workflow comparable to our setup; therefore, C++ bug-finding compares only \tool and Fuzz4All.

\begin{table}[htb!]
\centering
%\small
\setlength{\tabcolsep}{3.5pt}
\renewcommand{\arraystretch}{1.08}
\begin{tabular}{C{3cm}|C{1.3cm}|C{1.3cm}}
\toprule
\textbf{} & \textbf{Fuzz4All} & \textbf{\tool} \\
\midrule
\multicolumn{3}{c}{\textbf{G++ (C++ frontend)}} \\
\midrule
Unique & 12 & 24 \\
Not fixed in Trunk & 8 & 19 \\
\midrule
\multicolumn{3}{c}{\textbf{Clang++-23 (C++ frontend)}} \\
\midrule
Unique & 15 & 64 \\
\bottomrule
\end{tabular}
\caption{Unique compiler-internal failures over three independent 24-hour runs on C++ compilers. ``Not fixed in Trunk'' is reported only for G++.}
\label{tab:bugs_cpp_compilers}
\end{table}

%%\begin{table}[!htb]
  %  \centering
  %%  \small
   % \renewcommand{\arraystretch}{1.2}
   % \begin{tabular}{lccc}
   %     \toprule
   %     \textbf{-} & \textbf{Fuzz4All} & \textbf{FunFuzz} \\
    %    \midrule
     %   Unique & 12 & 24 \\
      %  Not fixed in Trunk & 8 & 19 \\
       % \bottomrule%
%    \end{tabular}
 %   \caption{Bug-finding results across three independent 24-hour runs on G++ (C++ frontend).}
  %  \label{tab:bugs_cpp}
%\end{table}
\textbf{G++}. Table~\ref{tab:bugs_cpp_compilers} shows that \tool finds substantially more unique C++ frontend failures than Fuzz4All (24 vs.\ 12). The same holds for persistence: 19 of these failures still reproduce on our pinned trunk snapshot, compared to 8 for Fuzz4All, indicating that \tool exposes a broader set of distinct and persistent failure modes.
%\begin{table}[!htb]
%    \centering
%    \small
%%    \renewcommand{\arraystretch}{1.2}
 %   \begin{tabular}{lccc}
  %      \toprule
   %     \textbf{-} & \textbf{Fuzz4All} & \textbf{FunFuzz} \\
    %    \midrule
     %   Unique & 15 & 64 \\
     %   \bottomrule
    %\end{tabular}
    %\caption{Bug-finding results across three independent 24-hour runs on Clang++-23 (C++ frontend).}
    %\label{tab:bugs_clang_cpp}
%\end{table}

\textbf{Clang++}. For \texttt{clang++-23} (trunk snapshot), we report only unique deduplicated failures. \tool again significantly outperforms Fuzz4All (64 vs.\ 15), suggesting that the bug-finding gains generalize to the richer compilation paths exercised by C++.

\subsubsection{Bug Diversity and Discovery Dynamics}
\label{sec:bug_diversity_dynamics}

To characterize bug diversity and discovery dynamics, we analyze (i) the overlap of deduplicated GCC failures across fuzzers and (ii) how new unique failures accumulate over time under a fixed 24-hour budget.

\begin{figure}[!htb]
    \centering
    \includegraphics[width=0.95\linewidth]{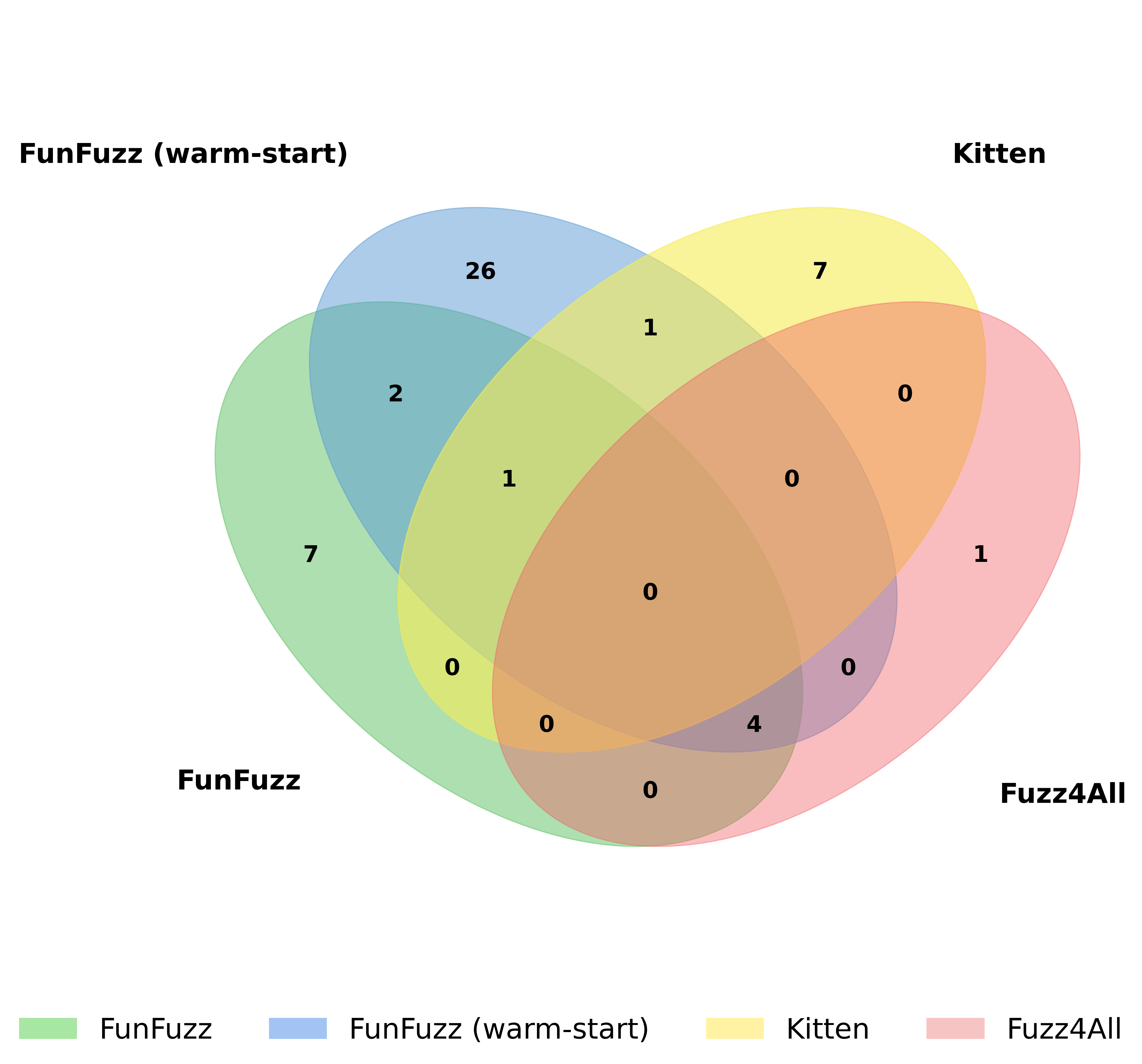}
    \caption{Overlap of unique GCC bugs discovered by different fuzzers.}
    \label{fig:venn_fuzzers}
\end{figure}

Figure~\ref{fig:venn_fuzzers} reports the set overlap of unique GCC failures (as defined by our crash-fingerprinting pipeline in Section~\ref{sec:crash_dedup}). \tool contributes a substantial fraction of failures that are not observed in the Fuzz4All or Kitten runs under the same budget, indicating that the corresponding inputs exercise failure modes that are not consistently reached by the baselines in our setting. Conversely, the number of failures exclusive to Fuzz4All is comparatively small. The warm-start configuration (\tool (warm-start)) yields the largest set of unique failures overall, suggesting that combining curated seeds with \tool's evolutionary loop increases the variety of reachable failure fingerprints beyond either component alone (while noting that this configuration is not baseline-aligned).

\begin{figure}[t]
    \centering
    \includegraphics[width=0.95\linewidth]{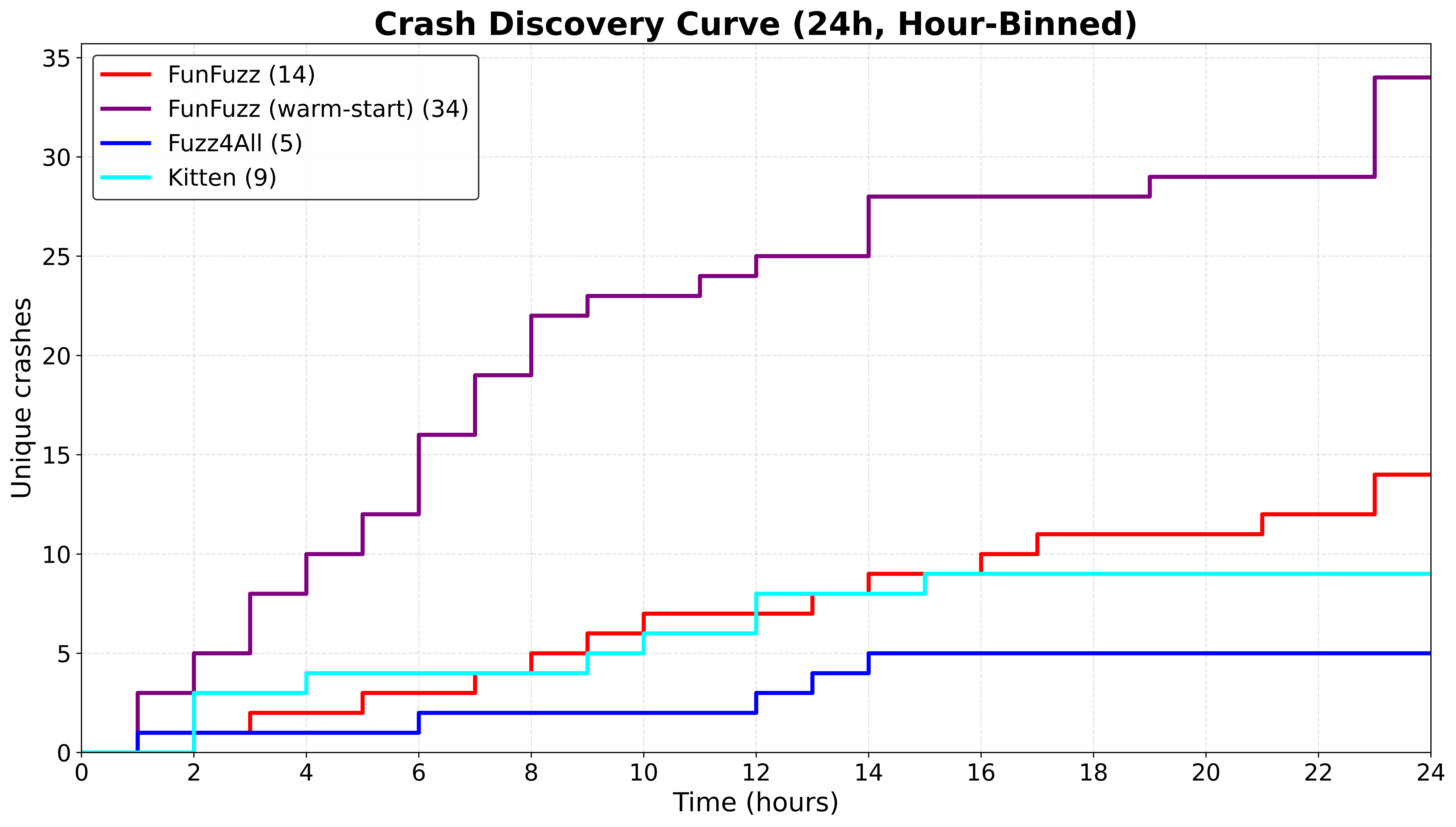}
    \caption{Cumulative unique GCC bugs discovered over time.}
    \label{fig:crash_discovery_curve_hourly_highres}
\end{figure}

Furthermore, Figure~\ref{fig:crash_discovery_curve_hourly_highres} plots the cumulative count of unique GCC failure fingerprints over time, where at each timestamp we take the union of newly observed fingerprints across the three runs. \tool continues to add new unique failures throughout the 24-hour campaign, whereas Fuzz4All saturates earlier and Kitten’s discovery rate slows later in the run. The warm-start configuration combines a fast early increase with sustained later discoveries, consistent with the overlap results.

\subsubsection{Triage Summary}
\label{sec:bug_triage}

We additionally track the triage and disclosure status of the failures found by \tool. Table~\ref{tab:bugs} summarizes a manual triage of the deduplicated failure fingerprints collected across our bug-finding campaigns (aggregated over \tool configurations). For each fingerprint, we attempt minimization and then report it (or match it) in the corresponding upstream tracker.

\begin{table}[htbp]
\centering
%\small
\setlength{\tabcolsep}{6pt}
\renewcommand{\arraystretch}{1.15}
\begin{tabular}{l|cc}
        \toprule
        \textbf{Status} & \textbf{GCC 16.0} & \textbf{Clang 23} \\
        \midrule
        Total failures & 68 & 51 \\
        \hline
        Confirmed      & 39 & 41 \\
        Duplicate      & 5 & 4 \\
        Pending        & 24 & 6  \\
        \bottomrule
\end{tabular}
\caption{Manual triage status of \tool's deduplicated failures aggregated across GCC-16.0 and Clang-23 campaigns. \textbf{Confirmed}: acknowledged by an upstream developer; \textbf{Duplicate}: matches an existing upstream issue; \textbf{Pending}: issue filed or linked but not yet confirmed. Counts reflect the status at the time of writing.}
\label{tab:bugs}
\end{table}

We label a fingerprint as Confirmed if the issue has been acknowledged as a real defect by an upstream developer on the official repository/tracker; as Duplicate if it matches an already-known defect (i.e., an existing upstream issue or an equivalent previously reported failure); and as Pending if we have filed or linked an upstream issue but it has not yet received developer confirmation or a definitive classification. These numbers should be interpreted as a snapshot of an ongoing reporting process rather than an exhaustive accounting. All reports are submitted through the official compiler issue trackers (e.g., LLVM’s \texttt{llvm-project} tracker, \texttt{GCC bugzilla}). For transparency and reproducibility, we provide in our public GitHub repository a complete list of all deduplicated failure fingerprints, together with their minimization status and links to the corresponding upstream issue IDs when available.

\subsection{Design Analysis and Ablation Study}
\label{sec:ablation}
We study which components of \tool drive coverage growth and bug-finding behavior by ablating individual design choices while keeping the rest of the pipeline fixed. Unless stated otherwise, we run short 30,000 programs campaigns on GCC (C) and g++ (C++) under the same generation model, decoding parameters, and evaluation harness as in the main experiments, and report compiler source-line coverage. These ablations are intended to characterize early search dynamics and motivate our default configuration, not to claim global optimality.

While the ablations in this section focus on coverage growth and search dynamics, \tool also supports feature-level steering via targeted prompting. We evaluate this capability separately in Appendix~\ref{append:targeted}, where targeted campaigns demonstrate high hit rates for specific C/C++ constructs without collapsing coverage. These results indicate that the evolutionary design analyzed here remains effective even under constrained, feature-biased generation.

We organize the analysis around three factors: (i) scoring and selection variants, (ii) diversity mechanisms (seed initialization and cross-island sharing), and (iii) the number of islands under a fixed generation budget.

\subsubsection{Scoring Function and Selection Strategy}
\label{sec:scoring_ablation}
We isolate the impact of individual scoring/selection components using single-factor ablations. Unless stated otherwise, we run 30,000-program campaigns across GNU (GCC/G++) and LLVM (Clang/Clang++) compilers, keeping the generation model, decoding parameters, and evaluation harness fixed. Table~\ref{tab:experiments_parameters_stacked} summarizes the variants; detailed definitions of each toggle are provided in Appendix~\ref{append:fitness}.

Across both infrastructure targets, incorporating feedback mechanisms consistently improves exploration. Incremental coverage scoring (\texttt{+Score}) yields substantial gains over the baseline (e.g., +25,933 lines in G++ and +47,888 in Clang). Adding compilation-time tracking (\texttt{+Time}) and redundancy filtering (\texttt{+Filter}) also strictly improves coverage in all environments by reducing repeated evaluations and favoring more structurally complex compiler paths. Notably, in the more permissive LLVM ecosystem, \textit{every} isolated modification—including failure rewarding and zero-filtering—injects useful diversity that uniformly and significantly improves coverage (+31k to +50k lines) over the baseline.

In sharp contrast, several knobs are counterproductive for GNU front-ends. Explicitly favoring compilation failures (\texttt{+Fail}) increases G++ coverage but severely reduces GCC coverage (-31,902 lines), suggesting failure-centric pressure misguides the search depending on the parser. Similarly, zero-score filtering (\texttt{+Zero}) improves G++ but actively hurts GCC (-11,151), while discarding used parents (\texttt{+Used}) is detrimental to both. Finally, although switching to a global coverage counter (\texttt{+Global}) yields strong early gains across all short ablations, we keep per-island accounting as the default to prevent premature cross-island convergence in 24-hour campaigns. Consequently, our unified configuration only enables globally robust features (\texttt{Score}, \texttt{Time}, and \texttt{Filter}).

\begin{table}[htbp]
\centering
\setlength{\tabcolsep}{4.5pt}
\renewcommand{\arraystretch}{1.08}

% --- Primera tabla (GCC) ---
\textbf{GNU Compilers}\\[1ex]
\begin{tabular}{C{1.3cm}|C{1.2cm}|C{1.3cm}|C{1.3cm}|C{1cm}|}
\toprule
\textbf{Variant} & \multicolumn{1}{c}{\textbf{G++}} & \multicolumn{1}{c}{\textbf{$\Delta$}} &
                   \multicolumn{1}{c}{\textbf{GCC}} & \multicolumn{1}{c}{\textbf{$\Delta$}} \\
\midrule
Baseline  & 243974 & --      & 237623 & -- \\
+Fail     & 252051 & +8077   & 205721 & -31902 \\
+Used     & 243035 & -939    & 221532 & -16091 \\
+Time     & 249220 & +5246   & 245252 & +7629 \\
+Score    & 269907 & +25933  & 243921 & +6298 \\
+Global   & 257523 & +13549  & 252726 & +15103 \\
+Zero     & 272498 & +28524  & 226472 & -11151 \\
+Filter   & 264239 & +20265  & 246648 & +9025 \\
\bottomrule
\end{tabular}

\vspace{0.2cm} % Espacio entre tablas

% --- Segunda tabla (Clang) ---
\textbf{LLVM Compilers}\\[1ex]
\begin{tabular}{C{1.3cm}|C{1.2cm}|C{1.3cm}|C{1.3cm}|C{1cm}|}
\toprule
\textbf{Variant} & \multicolumn{1}{c}{\textbf{Clang++}} & \multicolumn{1}{c}{\textbf{$\Delta$}} &
                   \multicolumn{1}{c}{\textbf{Clang}} & \multicolumn{1}{c}{\textbf{$\Delta$}} \\
\midrule
Baseline  & 653402 & --      & 535340 & -- \\
+Fail     & 687770 & +34368  & 568795 & +33455 \\
+Used     & 685445 & +32043  & 576580 & +41240 \\
+Time     & 684715 & +31313  & 582621 & +47281 \\
+Score    & 680973 & +27571  & 583228 & +47888 \\
+Global   & 686441 & +33039  & 586267 & +50927 \\
+Zero     & 684830 & +31428  & 582396 & +47056 \\
+Filter   & 686448 & +33046  & 578910 & +43570 \\
\bottomrule
\end{tabular}

\caption{Single-factor ablations of the scoring/selection pipeline for GNU and LLVM compilers (3-hour runs). Each row enables exactly one toggle relative to the baseline. Full definitions are in Appendix~\ref{append:fitness}.}
\label{tab:experiments_parameters_stacked}
\end{table}

\subsubsection{Seed Initialization, Sharing, and Diversity Control}
\label{sec:seed_sharing_ablation}

Beyond scoring, we ablate structural controls that shape early exploration: seed initialization, cross-island sharing, and parent selection. Unless stated otherwise, ablations evaluate 3-hour campaigns, while cross-island sharing is measured natively over 24 hours. Results for all environments are summarized in Table~\ref{tab:ablation_combined}.

\textbf{Seed Initialization.}
When all islands start from an identical seed, early exploration trajectories exhibit severe redundancies. Initializing each island with a distinct starting seed effectively distributes the search from the beginning, seamlessly improving early coverage across all environments. As shown in Table~\ref{tab:ablation_combined}, distinct seeds boost 3-hour coverage by 11.1\% in GCC and 7.8\% in G++, introducing smaller but steady benefits across the LLVM ecosystem.

\textbf{Cross-Island Sharing Strategy.}
We evaluate \tool's soft migration (which exchanges promising candidates globally without wiping the destination population) against a strict FunSearch-style policy that periodically resets entire islands. While full resets can occasionally inject drastic novelty, they discard valuable accumulated genetic context and destabilize the search. Over full 24-hour campaigns (Table~\ref{tab:ablation_combined}), maintaining continuous context via soft migration consistently attains higher final coverage than full resets for both C (+4.2\% in GCC, +1.3\% in Clang) and C++ (+19.8\% in G++, +1.1\% in Clang++).

\textbf{Coverage-Guided vs Random Selection.}
Finally, we compare sampling parents proportional to our coverage-based distribution against naive uniform random selection. In all environments, targeted coverage feedback decisively outperforms randomness (e.g., yielding 237,623 vs. 216,168 lines in GCC and 686,531 vs. 679,504 in Clang++; Table~\ref{tab:ablation_combined}). This emphasizes that fitness-proportionate selection is necessary to systematically drive the evolutionary loop toward productive compiler paths, rather than stalling out on low-yield or uncompilable mutants.

\begin{table}[htbp]
\centering
\setlength{\tabcolsep}{3pt}
\renewcommand{\arraystretch}{1.08}
\resizebox{\linewidth}{!}{
\begin{tabular}{l|rr|rr}
\toprule
& \multicolumn{2}{c|}{\textbf{GNU Compilers}} & \multicolumn{2}{c}{\textbf{LLVM Compilers}} \\
\cmidrule(lr){2-3} \cmidrule(lr){4-5}
\textbf{Setting/Variant} & \textbf{GCC (C)} & \textbf{G++ (C++)} & \textbf{Clang (C)} & \textbf{Clang++ (C++)} \\
\midrule
\multicolumn{5}{c}{\textbf{Seed initialization (3h)}} \\
\midrule
Same seed (shared init.) & 211\,257 & 245\,023 & 583\,888 & 683\,042 \\
Per-island seed & \textbf{234\,703} & \textbf{264\,093} & \textbf{591\,338} & \textbf{686\,531} \\
\midrule
\multicolumn{5}{c}{\textbf{Cross-island sharing (24h)}} \\
\midrule
Soft migration (no full reset) & \textbf{329\,905} & \textbf{334\,557} & \textbf{634\,091} & \textbf{727\,964} \\
Full reset (FunSearch-style)   & 316\,690 & 279\,186 & 625\,976 & 720\,183 \\
\midrule
\multicolumn{5}{c}{\textbf{Parent selection (3h)}} \\
\midrule
Coverage-guided & \textbf{237\,623} & \textbf{243\,974} & \textbf{591\,338} & \textbf{686\,531} \\
Random          & 216\,168 & 238\,510 & 569\,677 & 679\,504 \\
\bottomrule
\end{tabular}
}
\caption{Impact of diversity and sharing mechanisms across all compiler targets. Ablations run for 3 hours, except cross-island sharing strategies naturally evaluated over 24 hours.}
\label{tab:ablation_combined}
\end{table}

\subsubsection{Effect of the Multi-Island Architecture}
\label{sec:islands_ablation}

\tool uses a multi-island evolutionary architecture to maintain multiple partially independent search processes in parallel. We study how the number of islands affects coverage under a fixed 24-hour wall-clock budget.

In our setup, program generation is primarily driven by LLM inference throughput. As the number of islands increases, the overall generation rate quickly reaches a saturation point; therefore, additional islands do not necessarily increase the total number of evaluated programs, but mainly redistribute the available generation/evaluation budget across more parallel evolutionary trajectories (program counts per configuration are reported alongside the results).

\begin{figure}[htbp]
    \centering
    % Gráfico de C arriba
    \begin{subfigure}[b]{0.95\linewidth}
        \centering
        \includegraphics[width=\linewidth]{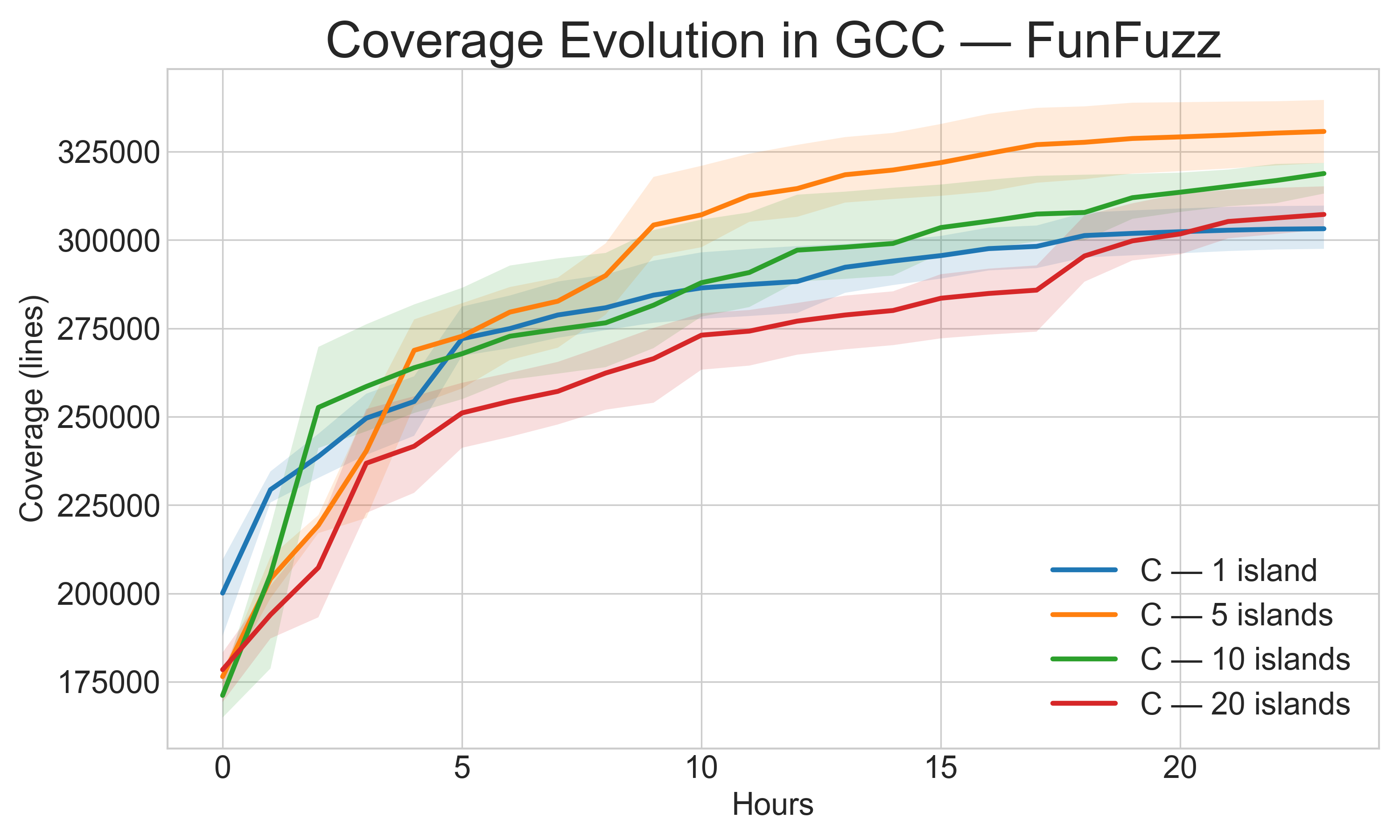}
        \caption{\textbf{C}}
        \label{fig:coverage_c_islands}
    \end{subfigure}
    
    \vspace{0.1cm} % Separación vertical para que respiren las gráficas
    
    % Gráfico de C++ abajo
    \begin{subfigure}[b]{0.95\linewidth}
        \centering
        \includegraphics[width=\linewidth]{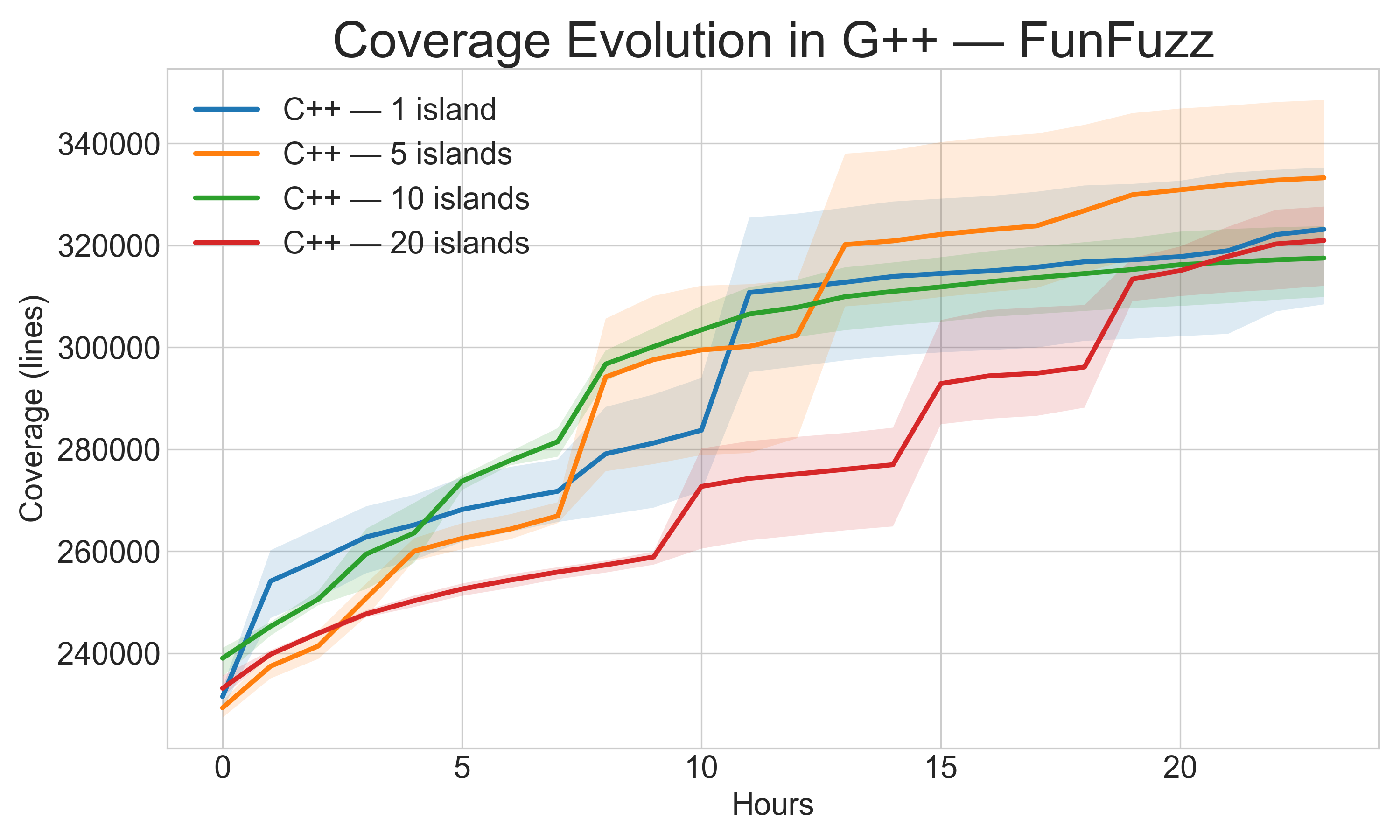} 
        \caption{\textbf{C++}}
        \label{fig:coverage_cpp_islands}
    \end{subfigure}
    
    \caption{Coverage evolution over 24 hours for different island configurations in \textbf{GCC} and \textbf{G++}.}
    \label{fig:coverage_islands} % Etiqueta general por si quieres referenciar a la figura entera
\end{figure}

Figures~\ref{fig:coverage_c_islands} and~\ref{fig:coverage_cpp_islands} show coverage over 24 hours for 1, 5, 10, and 20 islands in C and C++. Across both languages, we observe the same qualitative trend. A single island already achieves strong coverage growth, indicating that scoring and selection alone provide effective evolutionary pressure. Increasing to five islands yields the best overall performance, suggesting that moderate parallelism improves exploration by enabling divergent trajectories while preserving enough evolutionary depth within each island to refine promising directions.

In contrast, using 10 or 20 islands consistently underperforms under the same wall-clock budget. With the global input budget largely fixed, splitting it across too many islands reduces the number of updates per island and weakens selection pressure, leading to shallower evolutionary progress. In this regime, additional parallelism fragments the search rather than improving coverage.

%Overall, these results indicate that the benefits of island parallelism depend on the available generation capacity: moderate parallelism is advantageous under bounded throughput, while more aggressive parallelism would likely require higher global generation/evaluation rates to be beneficial.

\section{Conclusions} \label{sec:conclusions}
We presented \tool, an LLM-assisted evolutionary fuzzing framework that combines coverage-guided selection with a multi-island search procedure to sustain exploration on structured inputs. \tool initializes multiple semantically distinct search trajectories and uses lightweight migration to share high-value candidates without resetting populations. Across 24-hour campaigns on modern GCC and Clang for C and C++, \tool consistently improves compiler source-line coverage and discovers more unique compiler-internal failures than representative LLM and non-LLM baselines in our setting. Budget-aligned comparisons indicate that the gains are not explained solely by generating more inputs, but by producing higher-value programs. Ablations further show that coverage-based scoring is the main driver, while per-island initialization and conservative sharing help avoid premature convergence; moderate island parallelism provides the best trade-off under bounded generation throughput. Future work includes extending the oracle beyond internal failures to semantic bugs (e.g., miscompilations via differential testing), incorporating richer feedback signals (e.g., IR-level coverage or failure features) to refine fitness, and improving efficiency through hybrid mutation/generation and faster evaluation pipelines. We also plan to further systematize triage (minimization, clustering across versions, and mapping to upstream trackers) and to evaluate the approach on additional structured-input targets beyond compilers.

%% CCS: For help and more latex examples, refer to
%% `sample-sigconf.tex', provided in the distribution
%% https://portalparts.acm.org/hippo/latex_templates/acmart-primary.zip 
%%

%%
%% The acknowledgments section is defined using the "acks" environment
%% (and NOT an unnumbered section). This ensures the proper
%% identification of the section in the article metadata, and the
%% consistent spelling of the heading.

%% CCS: to preserve anonymity, NO acknowledgements to fundings, projects or persons should be used at
%% submission time
%% CCS: this section MAY be used to acknowledge the use of AI when used only for minor editorial improvements (e.g., grammar, spelling, or light style polishing) 
% \begin{acks}
% This paper was edited for grammar using [Tool Name].
% \end{acks}

%%
%% The next two lines define the bibliography style to be used, and
%% the bibliography file.
\bibliographystyle{ACM-Reference-Format}
\bibliography{biblio}

@article{mallissery2023demystify,
  title={Demystify the fuzzing methods: A comprehensive survey},
  author={Mallissery, Sanoop and Wu, Yu-Sung},
  journal={ACM Computing Surveys},
  volume={56},
  number={3},
  pages={1--38},
  year={2023},
  publisher={ACM New York, NY}
}

@article{ma2023survey,
  title={A survey of modern compiler fuzzing},
  author={Ma, Haoyang},
  journal={arXiv preprint arXiv:2306.06884},
  year={2023}
}

@article{kwon2025optimization,
  title={Optimization-Directed Compiler Fuzzing for Continuous Translation Validation},
  author={Kwon, Jaeseong and Jang, Bongjun and Lee, Juneyoung and Heo, Kihong},
  journal={Proceedings of the ACM on Programming Languages},
  volume={9},
  number={PLDI},
  pages={627--650},
  year={2025},
  publisher={ACM New York, NY, USA}
}

@misc{afl,
  author       = {Micha{\l} Zalewski},
  title        = {American Fuzzy Lop (AFL)},
  year         = {2014},
  howpublished = {\url{https://lcamtuf.coredump.cx/afl/}},
  note         = {Accessed: 2026-02-03}
}

@inproceedings{fioraldi2020afl++,
  title={$\{$AFL++$\}$: Combining incremental steps of fuzzing research},
  author={Fioraldi, Andrea and Maier, Dominik and Ei{\ss}feldt, Heiko and Heuse, Marc},
  booktitle={14th USENIX workshop on offensive technologies (WOOT 20)},
  year={2020}
}

@misc{libfuzzer,
  author       = {{LLVM Project}},
  title        = {libFuzzer -- a library for coverage-guided fuzz testing},
  howpublished = {\url{https://llvm.org/docs/LibFuzzer.html}},
  note         = {Accessed: 2026-02-03}
}

@inproceedings{xia2024fuzz4all,
  title={Fuzz4all: Universal fuzzing with large language models},
  author={Xia, Chunqiu Steven and Paltenghi, Matteo and Le Tian, Jia and Pradel, Michael and Zhang, Lingming},
  booktitle={Proceedings of the IEEE/ACM 46th International Conference on Software Engineering},
  pages={1--13},
  year={2024}
}

@article{romera2024mathematical,
  title={Mathematical discoveries from program search with large language models},
  author={Romera-Paredes, Bernardino and Barekatain, Mohammadamin and Novikov, Alexander and Balog, Matej and Kumar, M Pawan and Dupont, Emilien and Ruiz, Francisco JR and Ellenberg, Jordan S and Wang, Pengming and Fawzi, Omar and others},
  journal={Nature},
  volume={625},
  number={7995},
  pages={468--475},
  year={2024},
  publisher={Nature Publishing Group UK London}
}

@inproceedings{yang2011finding,
  title={Finding and understanding bugs in C compilers},
  author={Yang, Xuejun and Chen, Yang and Eide, Eric and Regehr, John},
  booktitle={Proceedings of the 32nd ACM SIGPLAN conference on Programming language design and implementation},
  pages={283--294},
  year={2011}
}

@inproceedings{holler2012fuzzing,
  title={Fuzzing with code fragments},
  author={Holler, Christian and Herzig, Kim and Zeller, Andreas},
  booktitle={21st USENIX Security Symposium (USENIX Security 12)},
  pages={445--458},
  year={2012}
}

@inproceedings{xie2025kitten,
  title={Kitten: A Simple Yet Effective Baseline for Evaluating LLM-Based Compiler Testing Techniques},
  author={Xie, Yuanmin and Xu, Zhenyang and Tian, Yongqiang and Zhou, Min and Zhou, Xintong and Sun, Chengnian},
  booktitle={Proceedings of the 34th ACM SIGSOFT International Symposium on Software Testing and Analysis},
  pages={21--25},
  year={2025}
}

@article{even2022csmithedge,
  title={CsmithEdge: more effective compiler testing by handling undefined behaviour less conservatively},
  author={Even-Mendoza, Karine and Cadar, Cristian and Donaldson, Alastair F},
  journal={Empirical Software Engineering},
  volume={27},
  number={6},
  pages={129},
  year={2022},
  publisher={Springer}
}

@inproceedings{deng2023large,
  title={Large language models are zero-shot fuzzers: Fuzzing deep-learning libraries via large language models},
  author={Deng, Yinlin and Xia, Chunqiu Steven and Peng, Haoran and Yang, Chenyuan and Zhang, Lingming},
  booktitle={Proceedings of the 32nd ACM SIGSOFT international symposium on software testing and analysis},
  pages={423--435},
  year={2023}
}

@article{huang2024large,
  title={Large language models based fuzzing techniques: A survey},
  author={Huang, Linghan and Zhao, Peizhou and Chen, Huaming and Ma, Lei},
  journal={arXiv e-prints},
  pages={arXiv--2402},
  year={2024}
}

@inproceedings{ou2024mutators,
  title={The mutators reloaded: Fuzzing compilers with large language model generated mutation operators},
  author={Ou, Xianfei and Li, Cong and Jiang, Yanyan and Xu, Chang},
  booktitle={Proceedings of the 29th ACM International Conference on Architectural Support for Programming Languages and Operating Systems, Volume 4},
  pages={298--312},
  year={2024}
}

@article{zhang2024llamafuzz,
  title={Llamafuzz: Large language model enhanced greybox fuzzing},
  author={Zhang, Hongxiang and Rong, Yuyang and He, Yifeng and Chen, Hao},
  journal={arXiv preprint arXiv:2406.07714},
  year={2024}
}

@inproceedings{eberlein2020evolutionary,
  title={Evolutionary grammar-based fuzzing},
  author={Eberlein, Martin and Noller, Yannic and Vogel, Thomas and Grunske, Lars},
  booktitle={International Symposium on Search Based Software Engineering},
  pages={105--120},
  year={2020},
  organization={Springer}
}

@article{li2019v,
  title={V-fuzz: Vulnerability-oriented evolutionary fuzzing},
  author={Li, Yuwei and Ji, Shouling and Lv, Chenyang and Chen, Yuan and Chen, Jianhai and Gu, Qinchen and Wu, Chunming},
  journal={arXiv preprint arXiv:1901.01142},
  year={2019}
}

@article{zhu2025software,
  title={When software security meets large language models: A survey},
  author={Zhu, Xiaogang and Zhou, Wei and Han, Qing-Long and Ma, Wanlun and Wen, Sheng and Xiang, Yang},
  journal={IEEE/CAA Journal of Automatica Sinica},
  volume={12},
  number={2},
  pages={317--334},
  year={2025},
  publisher={IEEE}
}

@inproceedings{huang2025challenges,
  title={On the challenges of fuzzing techniques via large language models},
  author={Huang, Linghan and Zhao, Peizhou and Ma, Lei and Chen, Huaming},
  booktitle={2025 IEEE International Conference on Software Services Engineering (SSE)},
  pages={162--171},
  year={2025},
  organization={IEEE}
}

@article{novikov2025alphaevolve,
  title={AlphaEvolve: A coding agent for scientific and algorithmic discovery},
  author={Novikov, Alexander and V{\~u}, Ng{\^a}n and Eisenberger, Marvin and Dupont, Emilien and Huang, Po-Sen and Wagner, Adam Zsolt and Shirobokov, Sergey and Kozlovskii, Borislav and Ruiz, Francisco JR and Mehrabian, Abbas and others},
  journal={arXiv preprint arXiv:2506.13131},
  year={2025}
}

@article{lange2025shinkaevolve,
  title={Shinkaevolve: Towards open-ended and sample-efficient program evolution},
  author={Lange, Robert Tjarko and Imajuku, Yuki and Cetin, Edoardo},
  journal={arXiv preprint arXiv:2509.19349},
  year={2025}
}

@inproceedings{herrera2021seed,
  title={Seed selection for successful fuzzing},
  author={Herrera, Adrian and Gunadi, Hendra and Magrath, Shane and Norrish, Michael and Payer, Mathias and Hosking, Antony L},
  booktitle={Proceedings of the 30th ACM SIGSOFT international symposium on software testing and analysis},
  pages={230--243},
  year={2021}
}

%%
%% Appendices
\appendix %% CCS: DO NOT REMOVE

\section{Open Science}
\label{append:open}
We provide an anonymized artifact repository containing the \tool implementation, configuration files, and scripts used to run the experiments and regenerate the key plots/tables in this paper: \url{https://anonymous.4open.science/r/paper-repository-6EE4/README.md}. The repository documents the required software dependencies and compiler builds, the prompt templates and initialization procedure, and the evaluation pipeline (coverage collection, failure detection, and crash deduplication). To support reproducibility, we also provide the exact experiment configurations used in the paper and a small set of sanity-check commands to validate the setup before running full campaigns. Where disclosure permits, we include aggregated experimental outputs needed to reproduce the reported numbers (e.g., coverage summaries and deduplicated failure fingerprints), and we maintain a mapping from confirmed failures to upstream issue IDs. The artifact is available at the anonymous repository linked below; licensing and usage terms are specified in the repository.

\section{Ethical Considerations}
Our work studies LLM-assisted compiler fuzzing, where the goal is to generate synthetic source programs that exercise diverse compiler behaviors and surface compiler-internal failures (e.g., crashes, ICEs, and assertion failures). This research is intended to improve the reliability and security of widely used language toolchains by helping developers identify and remediate defects earlier. The main stakeholders are compiler developers and maintainers (e.g., GCC/LLVM communities), researchers building testing infrastructure, software developers who rely on correct compilation, and downstream users affected by toolchain robustness. We expect predominantly defensive benefits: improved compiler robustness can reduce miscompilation risk, denial-of-service failures during builds, and latent weaknesses that may propagate through the software supply chain. Our experiments do not involve human subjects, personal data, private code, deception, or interaction with unsuspecting users; all inputs are machine-generated programs evaluated on public compiler codebases. The principal ethical concern is dual use: methods that improve bug-finding effectiveness could also be used to discover flaws for offensive purposes. We mitigate this risk by focusing on failure discovery rather than exploit development, avoiding operational weaponization detail, and following responsible disclosure practices for previously unknown issues via official upstream channels, including minimized reproducers when appropriate and coordinated publication timing when needed. We report aggregated, deduplicated outcomes and reference public tracker entries when already disclosed through upstream processes. Some residual dual-use risk remains unavoidable, but we judge that, under these mitigations and disclosure practices, the societal and security benefits of improving critical compiler infrastructure outweigh the remaining risks and justify publication.

\section{Prompt Distillation and Initialization: Illustrative Examples}
\label{append:autoprompt}
This appendix provides an illustrative example of the prompt distillation and initialization procedure described in Section \ref{sec:autoprompting}. The goal is to make explicit how raw documentation is transformed into structured prompts that guide the initial stages of the fuzzing process.

\subsection{Input Documentation and Initial Distillation}

The pipeline starts from raw textual input provided by the user. Depending on the target, this input may consist of compiler documentation, protocol specifications, or curated code excerpts.
Figure~\ref{fig:documentation} shows an excerpt of compiler-related
documentation used as input in our experiments.

\begin{figure}[!htb]
\centering
\begin{adjustbox}{max width=\linewidth}
\begin{tcolorbox}[
    colback=blue!3,
    colframe=blue!40!black,
    title=Documentation excerpt provided to the LLM,
    fonttitle=\bfseries,
    boxrule=0.3pt,
    arc=1mm,
    left=3mm,
    right=3mm,
    top=1.5mm,
    bottom=1.5mm,
    sharp corners,
    enhanced,
    breakable,
    width=\linewidth
]

\begin{lstlisting}[
basicstyle=\ttfamily\scriptsize,
breaklines=true,
columns=fullflexible,
frame=none
]
C Standard Library headers
The interface of C standard library is defined by the following collection of headers.

<assert.h>      Conditionally compiled macro that compares its argument to zero
<complex.h> (since C99) Complex number arithmetic
<ctype.h>       Functions to determine the type contained in character data
<errno.h>       Macros reporting error conditions
<fenv.h> (since C99)    Floating-point environment
<float.h>       Limits of floating-point types
<inttypes.h> (since C99)        Format conversion of integer types
<iso646.h> (since C95)  Alternative operator spellings
<limits.h>      Ranges of integer types
<locale.h>      Localization utilities
<math.h>        Common mathematics functions
<setjmp.h>      Nonlocal jumps
<signal.h>      Signal handling
<stdalign.h> (since C11)        alignas and alignof convenience macros
<stdarg.h>      Variable arguments
<stdatomic.h> (since C11)       Atomic operations
<stdbit.h> (since C23)  Macros to work with the byte and bit representations of types
<stdbool.h> (since C99) Macros for boolean type
...
<stdckdint.h> (since C23)       macros for performing checked integer arithmetic
\end{lstlisting}

\end{tcolorbox}
\end{adjustbox}
\caption{Example of external documentation injected into the LLM context.
Unlike the prompt template, this block represents factual specification
text used to guide semantic mutations.}
\label{fig:documentation}
\end{figure}

This raw input is forwarded to a  distillation LLM, which
generates a small set of candidate generic prompts summarizing the
documentation. Prompt generation follows a temperature-controlled sampling strategy designed to balance determinism and exploration. Specifically, one candidate is generated at temperature $T=0$ (conservative sampling), while three additional candidates are generated at temperature $T=1$ (exploratory sampling).

\begin{figure}[!htb]
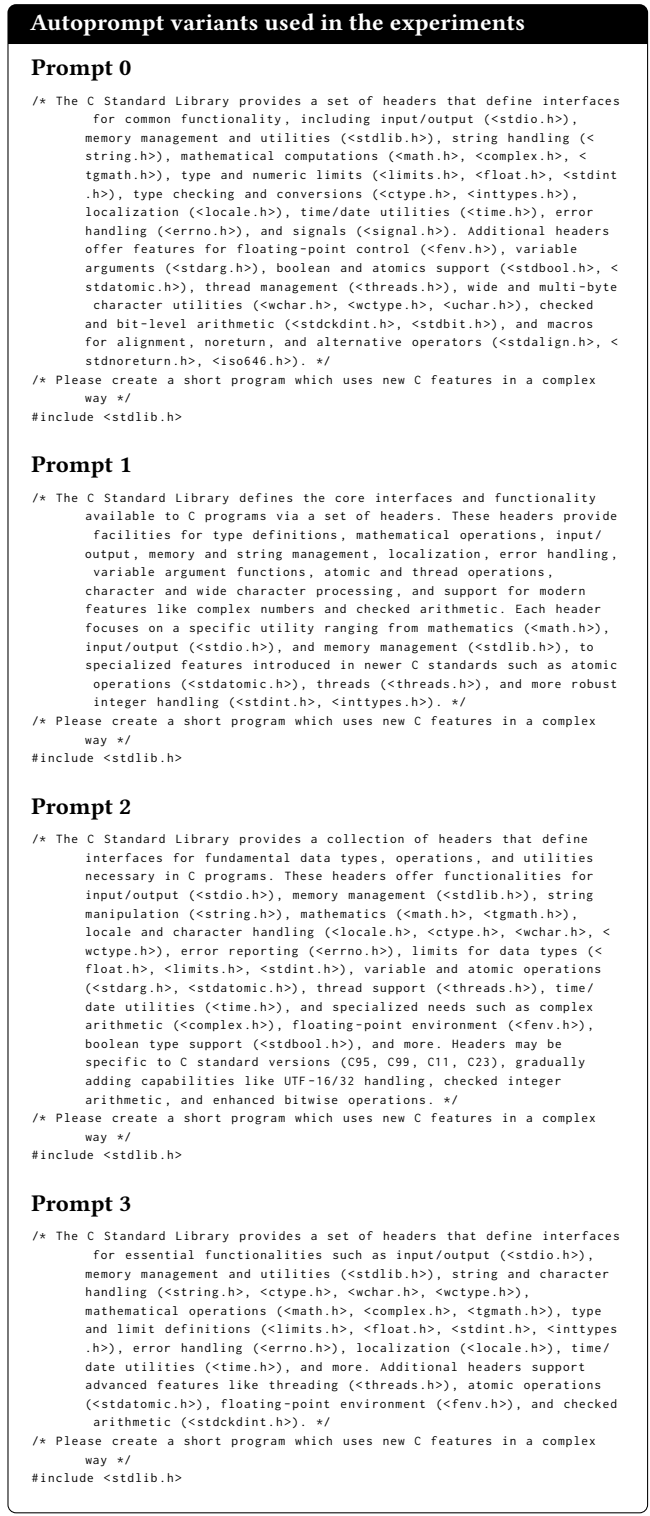

\centering
\begin{adjustbox}{max width=\linewidth}
\begin{tcolorbox}[
    colback=white,
    colframe=black,
    title=Autoprompt variants used in the experiments,
    fonttitle=\bfseries,
    boxrule=0.4pt,
    arc=1mm,
    left=2mm,
    right=2mm,
    top=1mm,
    bottom=1mm,
    enhanced,
    breakable,
    width=\linewidth
]

\textbf{Prompt 0}
\begin{lstlisting}[basicstyle=\ttfamily\tiny, breaklines=true]
/* The C Standard Library provides a set of headers that define interfaces for common functionality, including input/output (<stdio.h>), memory management and utilities (<stdlib.h>), string handling (<string.h>), mathematical computations (<math.h>, <complex.h>, <tgmath.h>), type and numeric limits (<limits.h>, <float.h>, <stdint.h>), type checking and conversions (<ctype.h>, <inttypes.h>), localization (<locale.h>), time/date utilities (<time.h>), error handling (<errno.h>), and signals (<signal.h>). Additional headers offer features for floating-point control (<fenv.h>), variable arguments (<stdarg.h>), boolean and atomics support (<stdbool.h>, <stdatomic.h>), thread management (<threads.h>), wide and multi-byte character utilities (<wchar.h>, <wctype.h>, <uchar.h>), checked and bit-level arithmetic (<stdckdint.h>, <stdbit.h>), and macros for alignment, noreturn, and alternative operators (<stdalign.h>, <stdnoreturn.h>, <iso646.h>). */
/* Please create a short program which uses new C features in a complex way */
#include <stdlib.h>
\end{lstlisting}

\vspace{0.5em}
\textbf{Prompt 1}
\begin{lstlisting}[basicstyle=\ttfamily\tiny, breaklines=true]
/* The C Standard Library defines the core interfaces and functionality available to C programs via a set of headers. These headers provide facilities for type definitions, mathematical operations, input/output, memory and string management, localization, error handling, variable argument functions, atomic and thread operations, character and wide character processing, and support for modern features like complex numbers and checked arithmetic. Each header focuses on a specific utility ranging from mathematics (<math.h>), input/output (<stdio.h>), and memory management (<stdlib.h>), to specialized features introduced in newer C standards such as atomic operations (<stdatomic.h>), threads (<threads.h>), and more robust integer handling (<stdint.h>, <inttypes.h>). */
/* Please create a short program which uses new C features in a complex way */
#include <stdlib.h>
\end{lstlisting}

\vspace{0.5em}
\textbf{Prompt 2}
\begin{lstlisting}[basicstyle=\ttfamily\tiny, breaklines=true]
/* The C Standard Library provides a collection of headers that define interfaces for fundamental data types, operations, and utilities necessary in C programs. These headers offer functionalities for input/output (<stdio.h>), memory management (<stdlib.h>), string manipulation (<string.h>), mathematics (<math.h>, <tgmath.h>), locale and character handling (<locale.h>, <ctype.h>, <wchar.h>, <wctype.h>), error reporting (<errno.h>), limits for data types (<float.h>, <limits.h>, <stdint.h>), variable and atomic operations (<stdarg.h>, <stdatomic.h>), thread support (<threads.h>), time/date utilities (<time.h>), and specialized needs such as complex arithmetic (<complex.h>), floating-point environment (<fenv.h>), boolean type support (<stdbool.h>), and more. Headers may be specific to C standard versions (C95, C99, C11, C23), gradually adding capabilities like UTF-16/32 handling, checked integer arithmetic, and enhanced bitwise operations. */
/* Please create a short program which uses new C features in a complex way */
#include <stdlib.h>
\end{lstlisting}

\vspace{0.5em}
\textbf{Prompt 3}
\begin{lstlisting}[basicstyle=\ttfamily\tiny, breaklines=true]
/* The C Standard Library provides a set of headers that define interfaces for essential functionalities such as input/output (<stdio.h>), memory management and utilities (<stdlib.h>), string and character handling (<string.h>, <ctype.h>, <wchar.h>, <wctype.h>), mathematical operations (<math.h>, <complex.h>, <tgmath.h>), type and limit definitions (<limits.h>, <float.h>, <stdint.h>, <inttypes.h>), error handling (<errno.h>), localization (<locale.h>), time/date utilities (<time.h>), and more. Additional headers support advanced features like threading (<threads.h>), atomic operations (<stdatomic.h>), floating-point environment (<fenv.h>), and checked arithmetic (<stdckdint.h>). */
/* Please create a short program which uses new C features in a complex way */
#include <stdlib.h>
\end{lstlisting}

\end{tcolorbox}
\end{adjustbox}

\caption{Complete autoprompting variants provided to the LLM.
Each prompt injects different documentation context while preserving
the same generation instruction.}
\label{fig:appendix_prompts}
\end{figure}

These prompts are not yet tied to a specific island or exploration strategy; instead, they act as global semantic summaries of the target specification.

\subsection{Validity-Based Prompt Selection}
Before entering the evolutionary fuzzing loop, \tool performs a lightweight \emph{prompt selection} phase to identify a single high-quality generic prompt to be used as the distilled base prompt in subsequent experiments. In this step, the unit of evaluation is the generation prompt (not individual programs).

For each candidate generic prompt (Figure~\ref{fig:appendix_prompts}), we sample a fixed batch of 90 programs using the same generation LLM and decoding settings as in our experiments, and compile each program under the same compiler configuration and timeout. We then compute the prompt's \emph{compile-validity score}, defined as the number
of generated programs that successfully compile without requiring any manual or automated repair. This metric captures how reliably a prompt steers the model toward syntactically and semantically well-formed programs, independent of downstream evolutionary effects.

\begin{table}[h]
\centering
%\small
\begin{tabular}{l c}
\toprule
\textbf{Prompt} & \textbf{compile-validity score} \\
\midrule
Prompt 0 & 27 \\
Prompt 1 & 20 \\
Prompt 2 & \textbf{30} \\
Prompt 3 & 17 \\
\bottomrule
\end{tabular}
\caption{compile-validity scores obtained during prompt selection. Each score corresponds
to the number of successfully compiling programs generated from a given prompt.
The highest-scoring prompt is selected as the distilled base prompt.}
\label{tab:validity_scores}
\end{table}

Table~\ref{tab:validity_scores}, reports the compile-validity results. \textbf{Prompt 2} achieves the highest value (30/90, 33.3\%) and is therefore selected as the shared distilled base prompt used in all subsequent fuzzing stages. The remaining candidates are discarded after this selection step.

\subsection{Island-Specific Seed Generation}

To encourage early semantic divergence across islands, \tool generates a set of island seed instructions from the same documentation used during the prompt distillation phase. These instructions are short, high-level directives—rather than concrete seed programs—designed to bias the LLM toward distinct program themes and feature combinations, such as control flow patterns, numeric corner cases, error handling, concurrency, or specific library usage.

\begin{figure}[!htb]
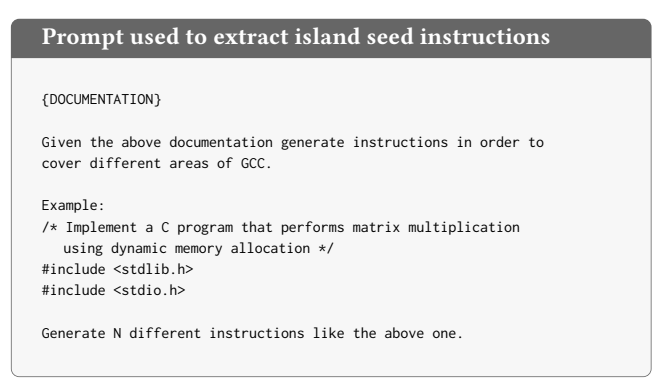

\centering

\begin{tcolorbox}[
    colback=black!3,
    colframe=black!60,
    boxrule=0.3pt,
    arc=1mm,
    left=3mm,
    right=3mm,
    top=2mm,
    bottom=2mm,
    enhanced,
    breakable,
    title=Prompt used to extract island seed instructions,
    fonttitle=\bfseries
]

\begin{lstlisting}[
basicstyle=\ttfamily\scriptsize,
breaklines=true,
columns=fullflexible
]
{DOCUMENTATION}

Given the above documentation generate instructions in order to
cover different areas of GCC.

Example:
/* Implement a C program that performs matrix multiplication
   using dynamic memory allocation */
#include <stdlib.h>
#include <stdio.h>

Generate N different instructions like the above one.
\end{lstlisting}

\end{tcolorbox}

\caption{Prompt used to generate candidate seed instructions for initializing island-specific corpora}
\label{fig:prompt_instructions}
\end{figure}

Concretely, we query the instruction-generation model using the raw documentation concatenated with a fixed instruction-generation prompt, illustrated in Figure~\ref{fig:prompt_instructions}. To balance determinism and exploration, two batches of instructions are generated:
\begin{itemize}
    \item a conservative batch sampled at temperature $T=0$,
    \item an exploratory batch sampled at temperature $T=1$.
\end{itemize}

\begin{figure}[!htb]
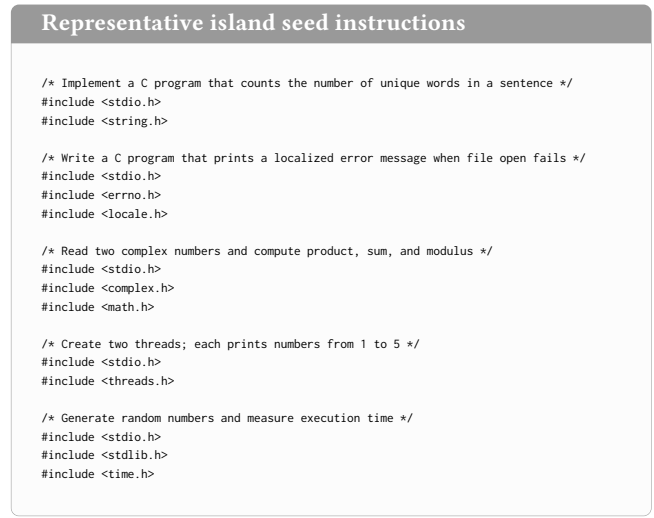

\centering

\begin{tcolorbox}[
    colback=black!1,
    colframe=black!40,
    boxrule=0.2pt,
    arc=1mm,
    left=3mm,
    right=3mm,
    top=2mm,
    bottom=2mm,
    enhanced,
    breakable,
    title=Representative island seed instructions,
    fonttitle=\bfseries
]

\begin{lstlisting}[
basicstyle=\ttfamily\tiny,
breaklines=true,
columns=fullflexible
]
/* Implement a C program that counts the number of unique words in a sentence */
#include <stdio.h>
#include <string.h>

/* Write a C program that prints a localized error message when file open fails */
#include <stdio.h>
#include <errno.h>
#include <locale.h>

/* Read two complex numbers and compute product, sum, and modulus */
#include <stdio.h>
#include <complex.h>
#include <math.h>

/* Create two threads; each prints numbers from 1 to 5 */
#include <stdio.h>
#include <threads.h>

/* Generate random numbers and measure execution time */
#include <stdio.h>
#include <stdlib.h>
#include <time.h>
\end{lstlisting}

\end{tcolorbox}

\caption{Example batch of island seed instructions generated by the LLM}
\label{fig:instrction_example}
\end{figure}

Each batch contains $N$ instructions, where $N$ corresponds to the number of islands. Representative examples of the generated island seed instructions are shown in Figure~\ref{fig:instrction_example}. These instructions are later combined with the distilled base prompt to construct the final hybrid prompts used during fuzzing.

Each island is assigned one instruction from the selected batch and evolves programs independently while preserving its assigned semantic focus.

\subsection{Hybrid Prompt Construction and Selection}

For each instruction batch generated in Section~A.3, hybrid prompts are
constructed by concatenating the distilled base prompt with the corresponding
island instructions. This produces two competing hybrid batches: one derived
from the $T=0$ seeds and another from the $T=1$ seeds (Figure
\ref{fig:prompt_complete}).

\begin{figure}[!htb]
    \centering
    \begin{adjustbox}{max width=\linewidth}
    \begin{tcolorbox}[
        colback=gray!5,
        colframe=black,
        title=Autoprompting final prompt,
        fonttitle=\bfseries,
        boxrule=0.4pt,
        arc=2mm,
        left=2mm,
        right=2mm,
        top=1mm,
        bottom=1mm,
        sharp corners,
        enhanced jigsaw,
        breakable,
        width=\linewidth
    ]

\textbf{Generic Prompt (template selected)}  
\begin{lstlisting}[language=C++, basicstyle=\ttfamily\footnotesize, breaklines=true]
/*
The C Standard Library provides a collection of headers that define interfaces for fundamental data types, operations, and utilities necessary in C programs. These headers offer functionalities for input/output (<stdio.h>), memory management (<stdlib.h>), string manipulation (<string.h>), mathematics (<math.h>, <tgmath.h>), locale and character handling (<locale.h>, <ctype.h>, <wchar.h>, <wctype.h>), error reporting (<errno.h>), limits for data types (<float.h>, <limits.h>, <stdint.h>), variable and atomic operations (<stdarg.h>, <stdatomic.h>), thread support (<threads.h>), time/date utilities (<time.h>), and specialized needs such as complex arithmetic (<complex.h>), floating-point environment (<fenv.h>), boolean type support (<stdbool.h>), and more. Headers may be specific to C standard versions (C95, C99, C11, C23), gradually adding capabilities like UTF-16/32 handling, checked integer arithmetic, and enhanced bitwise operations.
*/
\end{lstlisting}

\vspace{0.5em}
\textbf{Instruction Selected (semantic variant)}  
\begin{lstlisting}[language=C++, basicstyle=\ttfamily\footnotesize, breaklines=true]
/* Read two complex numbers and compute product, sum, and modulus */
#include <stdio.h>
#include <complex.h>
#include <math.h>
\end{lstlisting}

    \end{tcolorbox}
    \end{adjustbox}
    \caption{Prompt synthesized during the autoprompting phase for a given island. 
    The upper block is a reusable template inspired by Fuzz4All. 
    The bottom instruction is island dependent, to promote local diversity.}
    \label{fig:prompt_complete}
\end{figure}

For each hybrid prompt, a total of 60 programs are synthesized. Programs synthesized from all hybrid prompts generated under the same temperature are evaluated using the validity scoring pipeline defined in Section~A.2 The resulting validity scores are aggregated across all islands, producing a single global score per temperature configuration.

The global score obtained from the $T=0$ configuration is compared against the global score from the $T=1$ configuration. The higher-scoring configuration is selected in its entirety, and its hybrid prompts become the initial generation prompts assigned to the islands.

This set-level selection preserves semantic coherence across islands while still allowing controlled stochastic exploration through temperature sampling.

\section{Algorithm summary}
\label{append:algorithm}
\begin{algorithm}
\caption{Worker-driven fuzzing loop}
\label{alg:fuzzing_loop}
\begin{algorithmic}[1]
\Function{FuzzingLoop}{inputPrompt, timeBudget}
  \State \textbf{Input:} inputPrompt, timeBudget
  \State \textbf{Output:} bugs
  \State genStrats $\gets$ [generate-new, mutate-existing, semantic-equiv]
  \While{timeElapsed $<$ timeBudget}
    \State islandId $\gets$ random(nIslands)
    \State example $\gets$ getPromisingExample(islandId)
    \State instruction $\gets$ sample(genStrats)
    \State fuzzingInputs $\gets \mathcal{G}$(inputPrompt + example + instruction)
    \State (fitnessInputs, deltaCoverage) $\gets$ getMetric(fuzzingInputs, islandId)
    \State updateIsland(islandId, fuzzingInputs, fitnessInputs, deltaCoverage)
    \State checkIslandSharing(currentTime, lastSharing)
    \State bugs $\gets$ bugs + Oracle(fuzzingInputs, SUT)
  \EndWhile
  \State \Return bugs
\EndFunction
\end{algorithmic}
\end{algorithm}

\noindent

To illustrate the core logic of our fuzzing framework, Algorithm~\ref{alg:fuzzing_loop} presents a high-level description of the main control loop executed by a single worker. A worker is a lightweight orchestration entity responsible for driving the fuzzing process: it samples generation strategies, invokes the LLM to produce candidate inputs, dispatches these inputs to the evaluation pipeline, and updates the evolutionary state accordingly. Importantly, workers do not maintain evolutionary state themselves.

The evolutionary state is instead encapsulated within a set of independent \emph{islands}. Each island maintains its own local state, including its corpus of inputs, coverage information, fitness scores, and historical metadata. During each iteration, the worker selects an island, queries it for a promising example, and later updates the same island based on the evaluation results of newly generated inputs.

The fuzzing loop is instantiated independently by multiple workers, all operating on the same distilled \texttt{inputPrompt} but interacting with shared island states. This design enables concurrent exploration while preserving localized evolutionary dynamics within each island. Synchronization mechanisms are employed only when accessing shared island data, ensuring consistency without coupling the logical behavior of different workers.

\begin{itemize}
    \item \textbf{Line 4}: A predefined set of generation strategies (`genStrats`) is defined. These strategies encode different modes of generation:
    \begin{itemize}
        \item `generate-new`: produce novel code from scratch using only the distilled prompt.
        \item `mutate-existing`: modify previously successful examples.
        \item `semantic-equiv`: generate semantically similar variants of known programs.
    \end{itemize}

    \item \textbf{Lines 5–14}: The evolutionary loop runs until the time budget is exhausted.

    \item \textbf{Line 6}: A random island is selected, ensuring balanced usage over time.

    \item \textbf{Line 7}: The function `getPromisingExample(islandId)` retrieves a high-quality example program from the selected island, based on island-local fitness and novelty criteria.

    \item \textbf{Line 8}: A generation strategy is sampled from `genStrats`, determining the type of transformation applied during generation.

    \item \textbf{Line 9}: New fuzzing inputs are generated by invoking the LLM (`$\mathcal{G}$`) with a combination of the distilled prompt, the selected example, and the sampled instruction. Generation is thus implicitly conditioned on the semantic focus of the selected island.

    \item \textbf{Line 10}: Each generated program is executed and evaluated. The function `getMetric()` applies an island-specific fitness function, incorporating signals such as newly discovered compiler coverage, incremental scoring, and execution feedback. It returns per-input fitness values together with the incremental coverage contribution (`deltaCoverage`) of the current batch.

    \item \textbf{Line 11}: The island state is updated by integrating the newly evaluated programs, their fitness scores, and the associated incremental coverage, applying island-local selection and filtering policies.

    \item \textbf{Line 12}: The function `checkIslandSharing()` verifies whether the predefined sharing interval has elapsed and triggers island-level sharing or migration mechanisms if required.

    \item \textbf{Line 13}: Each batch of generated inputs is also evaluated by an external oracle (`Oracle()`), which detects crashes or anomalous behavior in the system under test (SUT). All detected failures are accumulated.
\end{itemize}

This high-level loop abstracts many of the architectural components described earlier (e.g., fitness-based selection, migration logic, filtering policies), providing a concise yet expressive view of the system’s behavior.
\section{Fitness Variants, Normalization, and Hyperparameters}
\label{append:fitness}
They include:

\begin{itemize}

\item \textbf{Failure Handling}: We evaluated whether programs that trigger compilation errors should be retained as examples in prompt construction. Although such programs are not directly executable, they may reflect boundary behaviors or syntax edge cases that contribute to discovering new coverage. Their inclusion thus influences the prompt update strategy and may help guide the LLM toward unexplored program behaviors.

\item \textbf{Reusing Code Snippets}: Repeatedly using successful programs in prompts may lead to premature convergence and reduced diversity. To mitigate this, we explored two strategies: (i) removing programs from memory once they are used in a prompt, and (ii) applying a score penalty to reduce their influence while keeping them available, the specified penalty was: \[
\text{new\_score} \;=\; \frac{\text{original\_score}}{10} \;-\; 1
\]
This design choice directly impacts the balance between retaining valuable examples and encouraging the emergence of novel behaviors.

\item \textbf{Time-Based Rewarding}: Compilation time may be used as a proxy for program complexity. We examined whether programs that take longer to compile should receive a proportional bonus to their fitness score, under the hypothesis that higher complexity may correlate with deeper structural features. This mechanism introduces a trade-off between rewarding structural richness and controlling evaluation overhead. The multiplier of the score is obtained following the formula:

\[
S = \frac{T_{\text{comp}}}{\overline{T}_{\text{prev}}} \times \max\!\left(8 - \overline{T}_{\text{prev}},\, 1\right)
\]

\noindent{Where:}
\begin{itemize}
    \item \(T_{\text{comp}}\) denotes the compilation time of the current program.
    \item \(\overline{T}_{\text{prev}}\) is the mean compilation time of the previously evaluated programs.
    \item The constant value \(8\) is a predefined hyperparameter chosen by us to stabilise the multiplier across different compilers. Since various compilers exhibit different baseline compilation times, subtracting \(\overline{T}_{\text{prev}}\) from this constant allows the multiplier to self-adjust depending on whether the compiler is relatively fast or slow.
\end{itemize}

\item \textbf{Incremental Coverage Scaling}: 
To reflect the increasing difficulty of discovering new behaviours as coverage grows, we apply a piecewise scaling scheme based on empirically observed coverage ranges. Let \(C_{\max}\) denote the maximum coverage value observed in historical executions of the fuzzer. This value is not fixed a priori; instead, it is continuously updated across campaigns as higher coverage levels are reached. In our current experimental setting, \(C_{\max}\) was approximately \(60{,}000\), based on the highest coverage observed during prior runs.

Using this empirical reference, we partition the reachable coverage space into three tiers:
\[
\text{Tier 1} = 0.40\,C_{\max}, 
\qquad 
\text{Tier 2} = 0.60\,C_{\max},
\]

\[
\text{Tier 3} = 0.80\,C_{\max}.
\]

These thresholds approximate early, intermediate, and late stages of exploration.

Each tier is associated with a maximum scaling multiplier:
\[
M_{1} = 5,\qquad M_{2} = 35,\qquad M_{3} = 100.
\]

Given the current coverage value \(C\), the scaling factor is selected according to the tier in which \(C\) falls. The rationale is the following: discovering new coverage early in the search is relatively easy, so the reward is modest, whereas finding coverage close to the frontier (i.e., above \(0.8\,C_{\max}\)) is significantly harder, and thus the multiplier is substantially larger. In contrast to prior descriptions that suggested a continuously increasing factor, our approach relies on discrete exploration stages, each capturing the increasing difficulty of extending coverage as fuzzing progresses.

\item \textbf{Coverage Accounting Mode}: We explored two scoring modes based on coverage tracking: (i) a global counter shared across all islands, and (ii) an independent counter per island (the default setting). The global mode encourages collaborative exploration by rewarding cross-island discoveries, while the independent mode promotes diversified behavior by allowing each island to evolve its own coverage frontier.

\item \textbf{Zero-Contribution Programs}: Programs that fail to increase coverage may still provide semantic value or structural variety. We evaluated policies for retaining these non-contributing samples as secondary resources, particularly under low-diversity conditions, where they might help enrich the prompt space and guide future generations.

\item \textbf{Redundancy Filtering}: To mitigate population stagnation, we implemented optional redundancy filters based on Levenshtein distance and Jaccard similarity. Programs exhibiting excessively high structural or lexical similarity were discarded, while those with moderate resemblance received a proportional penalty in their fitness score. This mechanism encourages diversity and reduces redundant computation.

\end{itemize}

\section{Reproducibility and Experimental details}
\label{append:experiments_configuration}

\subsection{Complete list of hyperparameters}

This appendix reports the complete set of experimental hyperparameters and implementation constants used across all experiments, summarized in Table~\ref{tab:hyperparams_defaults}.
 %It also details the compiler coverage instrumentation pipeline, normalization rules, and any parameters omitted from Table \ref{tab:exp_defaults_main} due to space constraints.
\begin{table*}[t]
\centering
%\small
\renewcommand{\arraystretch}{1.15}
\begin{tabular}{p{3.2cm} p{4.1cm} p{9.2cm}}
\toprule
\textbf{Stage} & \textbf{Parameter} & \textbf{Value} \\
\midrule

Prompt distillation & Distillation model &
GPT-4.1 (used only for prompt construction; not used for program sampling) \\

Prompt distillation & \# candidate prompts &
4 (1 sampled at $T=0$, 3 sampled at $T=1$) \\

Prompt distillation and Initiatilization & Init scoring signal &
Compile-validity: number of generated programs that compile successfully without manual intervention \\

Prompt distillation and Initiatilization & \# test programs per candidate prompt &
90 programs per prompt (validity-based evaluation) \\

Prompt distillation and Initiatilization & \# test programs per hybrid prompt &
60 programs per hybrid prompt \\

Prompt distillation and Initiatilization & Generation model & DeepSeek-Coder-V2-Lite-Base \\
Prompt distillation and Initiatilization & Decoding parameters &
Temperature $=1.0$; top-$p = 1.0$ (no truncation); max tokens $=512$; no top-$k$ or repetition penalties \\
Prompt distillation and Initiatilization & Samples per prompt ($n$) & 30 programs generated per prompt call \\
Prompt distillation and Initiatilization & Baseline alignment & \tool and Fuzz4All use identical decoding parameters for budget-aligned comparisons \\
Prompt distillation and Initiatilization & Inference backend &
vLLM API server (batched decoding enabled, bfloat16) \\

\midrule
Evolutionary fuzzing loop & Generation strategies & \texttt{generate-new}, \texttt{mutate-existing}, \texttt{semantic-equiv} \\
Evolutionary fuzzing loop & Island number & 5 (selected empirically; Section \ref{sec:islands_ablation})
 \\
Evolutionary fuzzing loop & Coverage accounting & Per-island independent coverage counters (default) \\
Evolutionary fuzzing loop & Coverage collection & Customized \texttt{libFuzzer} used for compiler-level coverage tracking \\
Evolutionary fuzzing loop & Program Fitness & Function described in Section \ref{sec:scoring_ablation}
 \\

\midrule
Cross-island migration & Migration period & Every 3 hours \\
Cross-island migration & Strong/weak split & Top 51\% islands = strong; bottom 49\% = weak \\
Cross-island migration & Weak-island pruning & Prune bottom 30\% of clusters in weak islands \\
Cross-island migration & Sharing rate & Each strong island shares 10\% of its population \\
Cross-island migration & Donor selection pool & Sample from the top-scoring 20\% subset of the strong island \\

\midrule
Evaluation protocol & Time budgets &
30000 programs (design/ablation); 24 hours (coverage + bug-finding); 10000 programs (Targeted)\\
Evaluation protocol & Repetitions & 3 runs with different random seeds (unless stated otherwise) \\

\bottomrule
\end{tabular}
\caption{Complete list of default \tool hyperparameters}
\label{tab:hyperparams_defaults}
\end{table*}

\subsection{Warm-start configuration}\label{appendix:hybridseed}
To assess whether \tool can benefit from curated compiler-specific inputs without abandoning its evolutionary prompting dynamics, we evaluate an additional ''warm-start`` configuration that mixes Kitten's corpus ( obtained from LLVM official suite) with internally generated candidates produced during the \tool campaign.

\textbf{Seed sources}. At each generation cycle, candidate programs are sampled from two sources:
\begin{itemize}
    \item External corpus: programs drawn from LLVM’s corpus for the target compiler.
    \item Internal pool: programs previously generated by \tool and stored in the program database together with their fitness scores.
\end{itemize}

\textbf{Mixing policy}. For each seed selection event, \tool samples from the external corpus with probability 0.5 and from the internal pool with probability 0.5. We use this 50/50 mixing ratio as a simple default to ensure that both sources contribute throughout the run; we do not tune this ratio for performance.

\textbf{Uniform treatment after sampling}. After selection, all seeds are treated identically by the evolutionary loop: they enter the same parent-selection and scoring pipeline, are compiled and scored under the local island’s fitness signals, and are subject to the same retention and pruning rules as any other candidate. No additional heuristics are applied to externally sourced seeds beyond their availability as parent candidates.

\textbf{Interpretation.} This configuration isolates the impact of target-specific initialization. By providing \tool with the same \textit{privileged information} used by Kitten---specifically, curated seeds sourced from open-access LLVM test suites---we can evaluate how our evolutionary loop performs in a \textit{warm-start} regime. These results serve as an upper-bound reference and demonstrate that \tool is complementary to existing corpora. To maintain a fair comparison with generative baselines, we distinguish these results from our standard ``from-scratch'' campaigns.

\section{Targeted Fuzzing Evaluation}
\label{append:targeted}
Following Fuzz4All's targeted-fuzzing setup (feature-specific documentation and hit-rate reporting) \cite{xia2024fuzz4all}, we assess \tool's ability to bias generation toward selected C/C++ constructs. Beyond general-purpose exploration, we evaluate whether \tool can steer generation toward specific language constructs while retaining broad compiler coverage. We run targeted campaigns for C (\texttt{typedef/union/goto}) and C++ (\textbf{std::apply} / \textbf{std::expected} / \textbf{std::variant}) by augmenting the user-provided documentation and instruction prompt with feature-specific guidance; the evolutionary loop (scoring, selection, and island management) is unchanged. Each campaign generates 10,000 programs. We report (i) hit rate, defined as the fraction of generated programs that contain the target construct according to our feature detector, and (ii) compiler source-line coverage accumulated during the campaign.

\begin{table}[htb!]
\centering
%\small
\caption{GCC targeted campaign}
\begin{tabular}{lcccc}
\toprule
\textbf{Hit rate} & \textbf{union} & \textbf{typedef} & \textbf{goto} & \textbf{General} \\
\midrule
union   &  80.66\% & 35\%       & 1.16\%     & 7.64\% \\
typedef  & 5.87\%      & 76.59\% & 0.20\%     & 13.6\%\\
goto   & 0.62\%      & 6.25\%     & 70.45\% & 1.22\% \\
\midrule
Coverage & 149671 & 158532 & 149340 & 182691 \\ 
\bottomrule
\end{tabular}
\label{tab:c}
\end{table}

\begin{table}[htb!]
\centering
%\small
\caption{G++ targeted campaign}
\begin{tabular}{lcccc}
\toprule
\textbf{Hit rate} & \textbf{apply} & \textbf{expected} & \textbf{variant} & \textbf{General} \\
\midrule
apply   &  67.26\% & 0.14\%       & 0.83\%     & 0.4\% \\
expected  & 0.19\%      & 79.63\% & 2.66\%     & 0.56\%\\
variant   & 0.52\%      & 0.46\%     & 85.73\% & 3.32\% \\
\midrule
Coverage & 213342 & 211625 & 210512 & 233846 \\ %186.742 186.302 191.121 195.512
\bottomrule
\end{tabular}
\label{tab:c++}
\end{table}

Tables~\ref{tab:c} and~\ref{tab:c++} show that targeted prompting yields consistently high hit rates for the intended constructs (e.g., 76.6\% for \textbf{typedef}, 80.7\% for \textbf{union}, and 85.7\% for \textbf{std::variant}), whereas general prompting produces substantially lower rates (ranging from 0.4\% to 13.6\% depending on the construct). Despite the additional constraints introduced by targeted guidance, compiler source-line coverage remains substantial across all campaigns, indicating that \tool does not collapse into repetitive or shallow instances of the target feature. Across all targeted campaigns, \tool triggers 15 deduplicated compiler-internal failures. Overall, these results show that targeted prompting remains effective even under a limited budget of 10{,}000 generated programs, enabling both construct-specific bias and the discovery of compiler-internal failures.

\end{document}